\documentclass[aps,pre,showpacs,superscriptaddress,final,floatfix]{revtex4-1}

\usepackage{amssymb}
\usepackage{amsmath}
\usepackage{color}
\usepackage{textcomp}
\usepackage[utf8]{inputenc}
\newcommand\ra{\rangle}
\newcommand\la{\langle}

\usepackage{graphicx}
\graphicspath {{./Figures/}}
\usepackage{textcomp}
\usepackage{amsmath}

\begin{document}

\title{Thermalization of local observables in the $\alpha$-FPUT chain}

\author{Santhosh Ganapa, Amit Apte and Abhishek Dhar}
\affiliation{International Centre for Theoretical Sciences, Tata Institute of Fundamental Research, Bengaluru -- 560089, India}

\begin{abstract}  Most studies on the problem of equilibration of the Fermi-Pasta-Ulam-Tsingou (FPUT) system have focused on equipartition of energy being attained amongst the normal modes of the corresponding harmonic system. In the present work, we instead discuss the equilibration problem in terms of local variables,  and consider initial conditions corresponding to spatially localized energy.
	We estimate the time-scales for equipartition of space localized degrees of freedom and find significant differences with the times scales observed for normal modes.  Measuring thermalization in classical systems necessarily requires some averaging, and this could involve one over initial conditions or over time or spatial averaging. Here we consider averaging over initial conditions chosen from a narrow distribution in phase space. We examine in detail the effect of the width of the initial phase space distribution, and of integrability and chaos,  on the time scales for thermalization. We show how thermalization properties of the system, quantified by its equilibration time, defined in this work, can be related to chaos, given by the maximal Lyapunov exponent.  Somewhat surprisingly we also find that the ensemble averaging can lead to thermalization of the integrable Toda chain, though on much longer time scales.

\vskip 0.5cm 
\textit{Dedicated to Joel Lebowitz, to thank him for being a constant source of new ideas, for explaining Boltzmann, and for  his warmth and kindness.}
\end{abstract}

\date{6 May 2020}

 \maketitle
\tableofcontents

 \section{Introduction}
\label{sec:Intro}
The Fermi-Pasta-Ulam-Tsingou (FPUT) problem \cite{Fermi1955,dauxois2008fermi} is a classic study with a long history. The main aim of the FPUT study was to observe thermal equilibration in an isolated  nonlinear system with Hamiltonian dynamics. A chain of oscillators with weak non-linear coupling was started in a highly atypical initial condition (all energy in a particular mode) and it was asked if the system evolves at long times to a state with equipartition of energy amongst different modes. To the surprise of the authors, they did not find such a state in the time scales of their observations. Instead they found quasi-periodic behaviour and near-recurrences to the initial state. 

Since the original work of FPUT, there have been a number of studies aimed at understanding their results, which led to significant developments  in statistical physics, nonlinear dynamics and mathematical physics. We only refer to several of the review articles on the FPUT problem \cite{Ford1992,Weissert1997,Berman2005,Gallavotti2008,Benettin2013}. Some of the recent studies that focus on the specific aspects of the FPUT problem that are especially relevant to this paper include the role of breathers \cite{Marin1996,Danieli2017,Flach2005,Flach2006,Christodoulidi2010}, wave-wave interaction theory \cite{Onorato2015,Lvov2018,Pistone2019}, and the role of breakup of invariant tori and the stochastic threshold \cite{Izrailev1966,Casetti1996,Deluca1995,Livi1985}, and these are described in detail in Sec.~\ref{sec:model}.

The main aim of the present study is to understand two specific aspects of the equipartition process: (i) the dependence on initial conditions, and (ii) the role of Lyapunov exponents, which themselves depend on the initial conditions for a Hamiltonian system. We are mainly motivated by the statement of the equipartition theorem and hence we focus on the averages $\left\langle z_i {\partial H}/{\partial z_i} \right\rangle$ where $z_i$ are phase space coordinates and $H({\bf z})$ is the Hamiltonian of the system.

The plan of the paper is as follows. In Sec.~\ref{sec:model} we define the precise model, and describe in some more detail the results of \cite{Onorato2015}, along with some of the other recent works relevant to this paper. We 
motivate and define the methods and objects of our study in Sec.~\ref{sec:methods}. In Sec.~\ref{sec:numerics} we present our numerical results  and then in Sec.~\ref{sec:chaos} we discuss the relation between these results to chaos. We conclude with a discussion in Sec.~\ref{sec:conclusions}.

In the rest of the introduction, we summarise some of the distinctive features of the approach we take.

(i) Dependence of equilibration time, $\tau_{\rm eq}$, on initial conditions and observed variables -- while most studies look at initial conditions specified in normal mode space, we study initial conditions with energy localized in real space. Similarly, most studies, beginning with the first FPUT paper have looked at equipartition of 
energy among normal modes. Here we  consider equipartition of local variables, which would be especially relevant when we consider strong nonlinearity, for which the normal mode picture becomes invalid. Several earlier works have discussed this in the context of harmonic chains \cite{mazur1960,titulaer1973a,titulaer1973b}.  

(ii) A necessary step required in studying the question of equilibration is that one needs some averaging process. Some commonly used protocols involve either an averaging over initial conditions (chosen from a specified initial distribution) \cite{Onorato2015} or a temporal averaging \cite{Benettin2013,Danieli2017,Benettin2018} or a spatial averaging \cite{Lebowitz1999,Goldstein2017}. Here we focus (see Sec.~\ref{ssec:avg-proc} for details) on  an averaging over initial conditions, and 
investigate the question of dependence of  the equilibration process on the width of the initial distribution.

(iii) Finally we explore the role of the Lyapunov exponents of the orbits in the process of equilibration. For a non-linear system, it is expected that a perturbation of the initial condition will grow exponentially for a non-integrable chaotic system and linearly for an integrable system. Hence we try to quantify in a precise way how a distribution of initial conditions expands over the full phase space and at sufficiently long times will reproduce the properties of the microcanonical equilibrium ensemble in not only chaotic systems but also anharmonic integrable models.

\section{Model and review of earlier work}
\label{sec:model}
We consider $N$ particles on a ring with displacements and momenta given by $\{q_i,p_i\}$, for $i=0,1,\ldots,N-1$. The system is governed by the following Hamiltonian:
\begin{equation}\label{eq:ham}
H(\boldsymbol{p},\boldsymbol{q})=\sum_{i=0}^{N-1}\left[\frac{p_i^2}{2m}+\frac{\mu (q_{i+1}-q_{i})^2}{2}+\frac{\alpha(q_{i+1}-q_{i})^3}{3}\right]~, 
\end{equation}
where we assume periodic boundary conditions $q_N \equiv q_0$ and $q_{-1} \equiv q_{N-1}$, and throughout the paper we use $m = \mu = 1$. 
The equations of motion are given by: 
\begin{equation} \label{eq:eqm}
\ddot{q}_i = (q_{i+1}+q_{i-1}-2q_{i})+\alpha \Big[(q_{i+1}-q_{i})^2-(q_{i}-q_{i-1})^2  \Big], 
\end{equation}
for $i=0,1,\ldots,N-1$. For $\alpha = 0$, we have a harmonic chain and we can make a linear change of variables to get an uncoupled set of $N$ oscillators.  For periodic boundary conditions, we define the Fourier modes
\begin{equation}\label{eq:nm}
Q_k =\frac{1}{\sqrt N} \sum_{j=0}^{N-1}q_je^{-i2\pi kj/N},~~ P_k =\frac{1}{\sqrt N} \sum_{j=0}^{N-1}p_je^{-i2\pi kj/N}.  
\end{equation}
The  Hamiltonian with $\alpha=0$ then takes the form
\begin{align}
H=\sum_{k=0}^{N-1} E_k,~~{\rm where}~ E_k=\Bigg[\frac{|P_k|^2}{2}+\frac{\Omega_k^2 |Q_{k}|^2}{2}\Bigg]~ \label{HNM}
\end{align}
is the energy of each mode and $\Omega_k = 2 \sin(k\pi/N), k = 0,1,2,...,N-1$. These are the normal modes of the system and since they do not interact, there is no exchange of energy  between them. The mode $k=0$ corresponds to the motion of the centre of mass of the system and we will always consider the case with $E_{k=0}=0$ (total momentum zero). If  the system is let to evolve from an arbitrary initial condition  $\{q_j(0),p_j(0)\}$, the energy of each mode $E_k$ remains constant with time.  For $\alpha\neq0$, the normal modes of the harmonic chain start interacting with each other and there is sharing of energy between the modes and the expectation is that this leads to equipartition of energy. For sufficiently small nonlinearity, the energy contribution from the nonlinear part of the interaction potential should be small and it is a good approximation to assume that the total energy can still be represented as a sum of the energies of the independent oscillators, i.e, the total energy $E \approx \sum E_k$. In that case, one check of equipartition would be to see if  all the $E_k(t)$ converge, at long times, to the same value $e=E/(N-1)$ (perhaps with small fluctuations). This was the approach in \cite{Fermi1955} (where  
however the fixed boundary condition case was studied). There, energy was initially given to the first normal mode ($E_1=0.08$) and with $\alpha=0.25$, the time evolution was studied numerically. Contrary to the expectations, the long-time dynamics  appeared to be almost periodic, with near perfect returns to the initial conditions. 

A heuristic estimate of the strength of the nonlinearity  can be obtained by comparing the contribution of the nonlinear interaction part to the total energy. 
Roughly, if $r$ is the average spacing between particles, we expect $\mu r^2\sim E/(N-1)=e$ which gives a length scale $r\sim\sqrt{e/\mu}$. An estimate of the ratio of the nonlinear and harmonic energies is then given by the parameter
\begin{equation}\label{eq:eps1}
\epsilon = \frac{\alpha r^3}{\mu r^2}=\frac{\alpha e^{1/2}}{\mu^{3/2}}~.
\end{equation}
This dimensionless number, $\epsilon$, and the system size, $N$, are the only relevant parameters. In our subsequent discussions we assume that $\mu=1$ and one can change $\epsilon=\alpha e^{1/2}$ by either changing the nonlinearity strength $\alpha$ or equivalently, by changing the energy density $e$.
Note that the cubic potential of the $\alpha$-FPUT system implies that the system stays bounded only if the total energy  is sufficiently small and the precise condition is $E< \mu^3/(6 \alpha^2)$, corresponding to all energy contributing to the potential energy of a single particle. In practice this is highly improbable and one can work with energies slightly higher than this bound.

We summarize some of the attempts to explain the absence of thermalization seen by FPUT.  

\begin{enumerate}
	\item \emph{Continuum limit and closeness to integrable PDEs}: Kruskal and Zabusky\cite{Zabusky1965} showed that the continuum limit of the model leads to the KdV equation which is an integrable model. One might then expect that the long wavelength initial condition used by FPUT remains close to the continuum description for long times and so one sees non-thermalization as expected for integrable systems. 
	
	\item \emph{Stochasticity threshold}: Izrailev and Chirikov \cite{Izrailev1966} proposed that there exists a  stochasticity threshold value of the energy density, $e_c$, such that for $e>e_c$ ($e=E/N$)  one gets thermalization. This idea was developed using the idea that nonlinearity leads to broadening of the normal mode frequencies and there is resonance overlap when the broadening is comparable to the separation between successive modes. The threshold depends on initial conditions and the system size. For the harmonic chain, for small values of $k(k \ll N$), the separation between successive levels scales as $\Delta_\omega \sim 1/N$, while for $N-k \ll N$, $\Delta_\omega \sim 1/N^2$.  
	Hence  the stochasticity threshold is larger for low frequency modes. Since the FPUT study had initial conditions with the lowest mode excited and the energy density was small, it is plausible that they were below the threshold.
	
	This idea has been studied numerically \cite{Casetti1996,Deluca1995,Livi1985} where an attempt was also made to relate this to chaotic properties. In particular it was pointed out that for generic initial conditions \cite{Casetti1996}, the time evolution of the maximum time-dependent Lyapunov exponent $\lambda(t)$, is indistinguishable for the $\alpha$-FPUT and the Toda systems up to a characteristic time $\tau_{\rm tr}$ (called the trapping time) that increase with decreasing energy (at fixed N). Beyond this time, $\lambda(t)$ of the $\alpha$-FPUT system appears to approach a constant $\Lambda$, while it keeps decreasing for the Toda system.
	This difference was attributed to the untrapping of the FPUT system from its regular region in phase space and escape to the chaotic component of its
	phase space. The authors in \cite{Casetti1996} computed $\tau_{\rm eq}, \tau_{\rm tr}$ and $\Lambda$ for system sizes $N=32,64,128$ at different energy densities and found the existence of a threshold $e_c(N)$ such that for $e < e_c$, $\tau_{\rm eq}, \tau_{\rm tr}$ seemed to diverge while $\Lambda$ vanishes. The threshold decreases with system size as $e_c(N) \sim 1/N^2$. Above $e_c$, power law dependences of the form $\tau_{\rm eq} \approx 1/e^3, \tau_{\rm tr} \approx 1/e^{2.5}$ and $\lambda \approx e^2$ were noted.  The FPUT parameters correspond to $e \ll e_c$.

	\item \emph{Role of breathers}: Breathers are time-periodic and space-localized solutions of nonlinear dynamical systems \cite{Marin1996}. By nature they are non-thermal and one might expect them to play a role in preventing thermalization.  In a certain sense the idea  is similar to the one relating the presence of solitons in the KdV system to the absence of equilibration in the FPUT system -- the difference being that breathers are stable solutions of the discrete system while the KdV is a continuum approximation. 
	The unexpected recurrences in the FPUT problem have been linked to the choice of initial conditions used by FPUT, which are set close to exact coherent time-periodic (or even quasiperiodic) trajectories, e.g., $q$-breathers, which show exponential localization of energy in normal mode space \cite{Danieli2017,Flach2005,Flach2006,Christodoulidi2010}. 
	
	\item  \emph{Ideas from wave turbulence}: The FPUT problem has recently been studied \cite{Onorato2015,Lvov2018,Pistone2019} using approaches of  wave turbulence \cite{Zakharov1992,Nazarenko2011}. Based on requirement of resonance between sets of normal modes, a detailed prediction has been made for the time-scale for equilibration and it is argued that this is finite for any non-zero strength of non-linearity, for finite sized systems. In particular, for the case of cubic nonlinearity of strength $\epsilon$ in dimensionless units, the estimated equilibration time scales as $\epsilon^{-8}$ for $N=16,32,64$.  In the thermodynamic limit, this  is predicted to change to the form $\epsilon^{-4}$.  The idea of the approach is to  connect the equilibration issue to the presence of high order resonances between dressed normal modes that appear under repeated canonical transformations.  The resonances are expected to lead to the irreversible transfer of energy and hence thermalization. Their work\cite{Onorato2015} showed that resonant triads (three wave resonances) are forbidden, they would generate a reversible dynamics, which was originally observed by Fermi. They looked for higher order interactions which are responsible for long term dynamics by a sequence of canonical transformations. They found that four wave resonant interactions, though they exist, are isolated from other quartets and cannot spread the energy across the spectrum. The six wave resonant interactions are interconnected and are the lowest order interactions that lead to an effective irreversible transfer of energy. In order to numerically verify the predicted equilibration time, the authors in \cite{Onorato2015} took  $N = 32$ particles on a ring, and considered two sets of initial conditions. In one set, energy was given only to one normal mode ($E_1, E_{31} \ne 0$), while in the other set, energy was given to five normal modes ($E_i \ne 0$ for $i=1,\dots,5$ and $i=28,\dots,32$).  Averages were taken over an ensemble of $1000$ initial conditions by introducing different random phases to each member. From the time evolution of the ensemble averaged normal mode energies, the time to achieve  equipartition was estimated and for the $\alpha$-FPUT model it was verified that the equilibration time scales as  $\sim 1/\epsilon^8$. A more recent study \cite{Lvov2018} of the $\beta$-FPUT chain suggests that wave-wave resonances lead to thermalization at small $\epsilon$, where $\tau_{\rm eq} \sim 1/\epsilon^4$, while at larger $\epsilon$ the level broadening mechanism leads to $\tau_{\rm eq} \sim 1/\epsilon$. It was suggested that no threshold exists. 
	
\end{enumerate} 

In most of the literature, the  analysis of equipartition was done for the energy of normal modes of the corresponding integrable problem, that is, the harmonic chain. The analysis is in some sense ``global,'' since normal modes involve all the particles in the chain. The present work attempts to analyse the problem ``locally,'' by checking equipartition theorem at different sites in the chain and tries to compute the time scale the system needs to reach equipartition. 
We also investigate the role of the initial ensemble in determining equilibration.

\section{Checking thermalization in an isolated system}
\label{sec:methods}

\subsection{Averaging procedures}
\label{ssec:avg-proc}
One can use various methods to check whether a system has reached thermal equilibrium. It is clear that to check equilibration during the time evolution requires some kind of averaging.  The procedure we follow here involves an 
averaging over initial conditions, with the expectation that such averages would represent the typical behaviour.  We perform an average over the initial conditions $({\bf q}_0,{\bf p}_0)$ which are now chosen from a narrow distribution centred around a specified point and that are still on the microcanonical surface of constant energy $E$, momentum $P$ and number of particles $N$. Denoting the initial distribution by $\rho_I({\bf q}_0,{\bf p}_0)$, we then obtain   the following average for any observable $A=A({\bf q},{\bf p})$:
\begin{equation}\label{eq:eavg}
\langle A \rangle(t)= \int d {\bf q}_0 d{\bf p}_0~ A({\bf q}(t),{\bf p}(t)) \rho_I({\bf q}_0,{\bf p}_0)~.  
\end{equation}
We then ask if this reproduces the expected equilibrium results which would be obtained from the equilibrium ensemble corresponding to the macroscopic conserved variables, e.g. total energy and number of particles. 
In particular we can ask if equipartition is achieved at long times, and the time to do so. 

In the present work, we focus on this protocol, to explore the issue of equilibration times in the FPUT problem. 
Some interesting questions include the dependence of the equilibration process 
on the ``width" of the initial distribution, and the fluctuations in the measured values. We  explore some of these questions.

{\bf Choice of initial distribution}: Here we consider the case where the initial distribution lies on the constant $(E,P,N)$ surface in $2N$ dimensional phase space and has a small spread, with the size of the spread characterized by  a dimensionless number $\gamma$. The initial condition is also chosen to correspond to the initial energy of the chain being localized in a small region.
Thus we set $q_i=0$, for $i=1,2,\ldots,N$ and $p_i=0$ for $i=5,6\ldots,N-1,N$. The total energy $E$ is then distributed amongst the four remaining particles in the following way
\begin{eqnarray}
E_1&=&E_2=(1-\gamma)E/4 +\nu \gamma E/2,~ \label{eq:ic1}\\
E_3&=&E_4=(1-\gamma)E/4 +(1-\nu) \gamma E/2,~\label{eq:ic2} 
\end{eqnarray}
where $0<\gamma <1 $ is a number which specifies the ``width" of the distribution and  $\nu$ is a uniformly distributed random number in the interval $(0,1)$ which basically gives us some randomness in the initial conditions. Here we consider initial conditions that have zero momentum. The first and the second particles are given velocities in opposite directions as are the third and the fourth.
In our simulations we generate $R$ initial configurations from this distribution and evolve the system with the Hamiltonian dynamics.  The ensemble average of a physical observable $A({\bf q},{\bf p})$ is estimated as 
$$ \langle A \rangle (t) = \frac{\sum_{r=1}^R A_r(t)}{R}, $$ where the sum is over the $R$ members of the ensemble.

Other averaging protocols commonly used in FPUT studies is temporal coarse graining. In this case one starts with a fixed initial condition and performs an average over time. In this case, one can look at microscopic variables, e.g the kinetic energy of individual particles, and ask whether the time averaged value corresponds to the expected equilibrium value. Starting the time evolution of the system from an arbitrary initial condition $({\bf q}_0,{\bf p}_0)$, with energy $E$, we can define a time averaged quantity for the observable $A=A({\bf q},{\bf p})$ as 
\begin{equation}\label{eq:tavg}
\overline{A}(t)= \frac{1}{t}\int_0^t ds A({\bf q}(s),{\bf p}(s))~.
\end{equation}
For a non-integrable system we might  expect that for generic initial conditions, at long times we should get thermal equilibration or 
$\overline{A}(t\to \infty) = \left\langle A\right\rangle_E$. The time to reach the equilibrium value should give a measure of equilibration time scales.
In some studies an average is taken over  finite large windows of time, centred at different time instances.

\subsection{Equipartition and estimation of  thermalization timescale}
\label{ssec:equipartition}
A standard test of thermalization would be to check the equipartition theorem expected in equilibrium systems. Let us recall the precise statement on equipartition. For a system in thermal equilibrium, the following are true:
\begin{equation}
\left\langle q_i\frac{\partial H}{\partial q_j} \right\rangle_{eq}=\left\langle p_i\frac{\partial H}{\partial p_j}\right\rangle_{eq}= c \delta_{ij},  \label{eqp1}
\end{equation}
for all $i,j=0,1,\ldots,N-1$ and where $\left\langle...\right\rangle_{eq}$ represents an  average over an equilibrium ensemble and  $c$ is a constant equal to $k_B(\partial S/\partial E)^{-1}$ for the microcanonical ensemble and equal to $k_BT$ for a canonical ensemble. 
In the rest of the paper we use the  notation 
\begin{align}
\left\langle T_i \right\rangle = \frac{1}{2} \left\langle p_i\frac{\partial H}{\partial p_i}\right\rangle \,, \quad \textrm{and} \quad \left\langle V_i \right\rangle =  \left\langle q_i\frac{\partial H}{\partial q_i} \right\rangle \,. \label{eq:avg-t-v}
\end{align}
For a weakly nonlinear system, we can neglect the nonlinear terms and then  transform to normal modes coordinates where the system looks like a collection of independent oscillators as in Eq.~\eqref{HNM}. In this case the equipartition theorem gives 
\begin{align}
\la E_k \ra_{eq} = E/N  \label{eqp2}
\end{align}
for the microcanonical ensemble and $k_B T$ for the canonical ensemble. We need to remove the zero mode if we are considering periodic boundary conditions. It is important to remember that Eq.~\eqref{eqp2} is an approximate form valid for weak nonlinearity while Eq.~\eqref{eqp1} is exact.

As a measure of the level of equipartition that is achieved, we define the  following  function \cite{Livi1985}, which has been referred to in the literature as entropy:

\begin{equation}\label{eq:ent}
S(t) = -\sum_{i=0}^{N-1} f_i(t) \ln f_i(t)
\end{equation}
where $ f_i(t) = \nu_i(t)/\sum_{r=0}^{N-1} \nu_r(t)$ for $i=0,1,\dots,N-1$, and $\nu_i$ could be either $\la T_i\ra$ or $\la V_i \ra$ or $\la E_i \ra$, which correspond to monitoring the equipartition of kinetic energy, $q_i\frac{\partial H}{\partial q_i}$, or the normal mode energy, respectively\footnote{Note that for each fixed time $t$, the set $\left\{ f_i(t) \right\}$ defines a discrete probability distribution over the set $\left\{0,1,\dots,N-1\right\}$, and then $S(t)$ is just the information entropy.}.  The value of $S(t)$ is bounded between $0$, corresponding to the highly nonequilibrium situation with all the energy in a single degree of  freedom, and $\ln N$, corresponding to the equilibrated system with equipartition between all degrees with $\left\{f_i\right\}$ defining a uniform distribution over the set $\left\{0,1,\dots,N-1\right\}$.
Since it can theoretically take an infinite amount of time to reach $\ln N$, we estimate the equilibration time as the time  required for $S(t)$ to reach a  predetermined value of entropy which is close to the equilibration value.
In this work, we consider the following criterion to determine the equilibration timescale:
\begin{equation}\label{criterion}
\left|\frac{S(t)-S_{max}}{S_{max}}\right| \leq 0.01~. 
\end{equation}
The above threshold must be satisfied for two consecutive values of the time that is sampled. The minimum value of $t$ for which the above criteria is satisfied is termed as equilibration time (denoted by $\tau_{\rm eq}$).

\section{Numerical results}
\label{sec:numerics}
As noted in the introduction, our main objectives are to estimate the equilibration time using observables other than the normal mode energies and to investigate the role of initial conditions. In the following, we present results on equilibration from the initial ensemble discussed in Sec.~\ref{ssec:avg-proc} corresponding to energy being initially localized in space. We shall refer to these initial conditions as space localized excitations (SLE) as opposed to normal mode localized excitations (NMLE) commonly used in most studies. Equipartition will be checked by monitoring the entropy $S(t)$ for the local observables   $\langle T_j \rangle$ and $\left\langle V_j\right\rangle$ which are defined in Eq.\eqref{eq:avg-t-v}. 
These quantities for the  $\alpha-$FPUT system are compared with those of the Toda chain and of the harmonic chain, to check the dependence of thermalization on integrability. This is quantified further by the analysis of the dependence of the entropy $S(t)$ on time $t$. In addition, the entropy of  $\left\langle V_j\right\rangle$ and $\left\langle T_j\right\rangle$ of the  $\alpha-$FPUT problem are compared with $\left\langle E_k\right\rangle$ to check if equilibration depends on whether one is verifying equipartition using the normal mode or position-momentum coordinates. The scaling laws in the two methods are compared by plotting the equilibration times as a function of the nonlinear dimensionless parameter $\epsilon$. The dependence of the equilibration time on the width of the initial distribution in the phase space is also analysed.

In Sec.~\ref{sec:chaos}  an attempt  is made to relate the thermalization properties of the $\alpha-$FPUT system (equilibration time) to the growth of perturbations of initial conditions in the system. For chaotic systems this growth is exponential and is  quantified by the maximal Lyapunov exponent, while for anharmonic integrable systems, the growth is linear. We make comparisons of the FPUT results with those of the corresponding Toda system. 

{\bf Simulation details}: For most of our numerics, we used a sixth order symplectic integrator described in \cite{Yoshida1990}. The time-step size was taken as $0.01$ and we checked the relative energy change in the system at the end of the computation to be of the  order of $10^{-11}$. Computations of the  Lyapunov exponent  were performed by solving the coupled system of $2N+2N$ nonlinear and linearized equations using a fourth order  Runge-Kutta integrator.  In this case the time-step size was taken as $0.001$ and  the relative energy change in the system at the end of the computation is found to be around $10^{-9}$.

\begin{figure}[ht]
	\centering
	\hspace{-10mm}
	\includegraphics[width=0.55\textwidth]{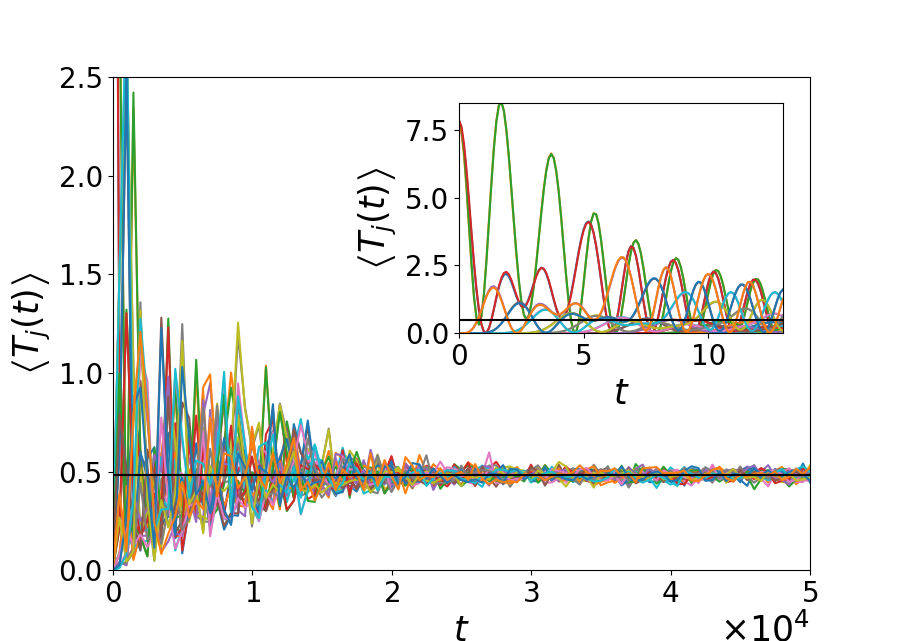}
	\put (-150,183) {$(a)$}
	\hspace{-10mm}
	\includegraphics[width=0.55\textwidth]{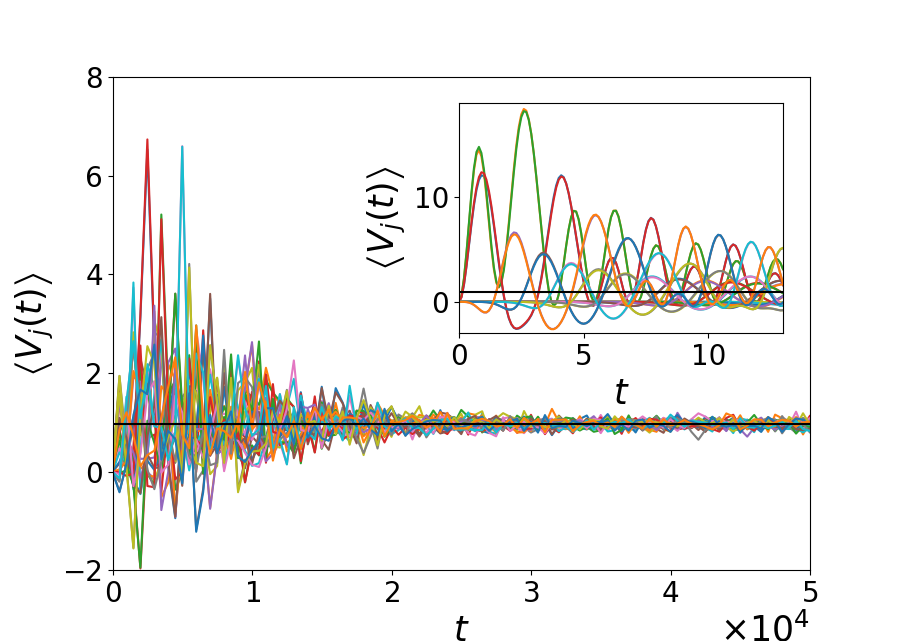}
	\put (-150,183) {$(b)$}
	\hspace{-10mm}
	\caption{$\alpha-$FPUT chain: Plot shows time evolution of $\left\langle T_j\right\rangle$ (left panel) and $\left\langle V_j\right\rangle$ (right panel) starting from space localized initial conditions for $j = 0,1,2,...N-1$.  Parameter values for this plot are $N=32$, $E=31$, $\alpha= 0.0848$ ($\epsilon=0.0834$), $\gamma=0.9$ and  $R=1000$. The solid black line corresponds to the equipartition value $\langle T_j \rangle= 0.484$ and $\langle V_j \rangle= 0.969$. The inset shows the time evolution at the earliest times.}
	\label{kvgp9}
\end{figure}
\begin{figure}[ht]
	\centering
	\hspace{-10mm}
	\includegraphics[width=0.55\textwidth]{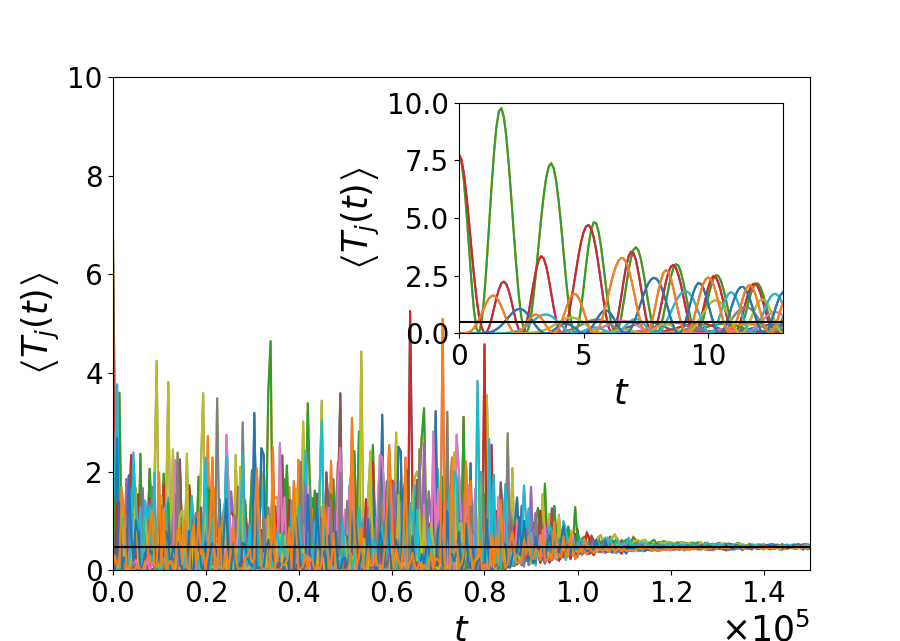}
	\put (-150,183) {$(a)$}
	\hspace{-10mm}
	\includegraphics[width=0.55\textwidth]{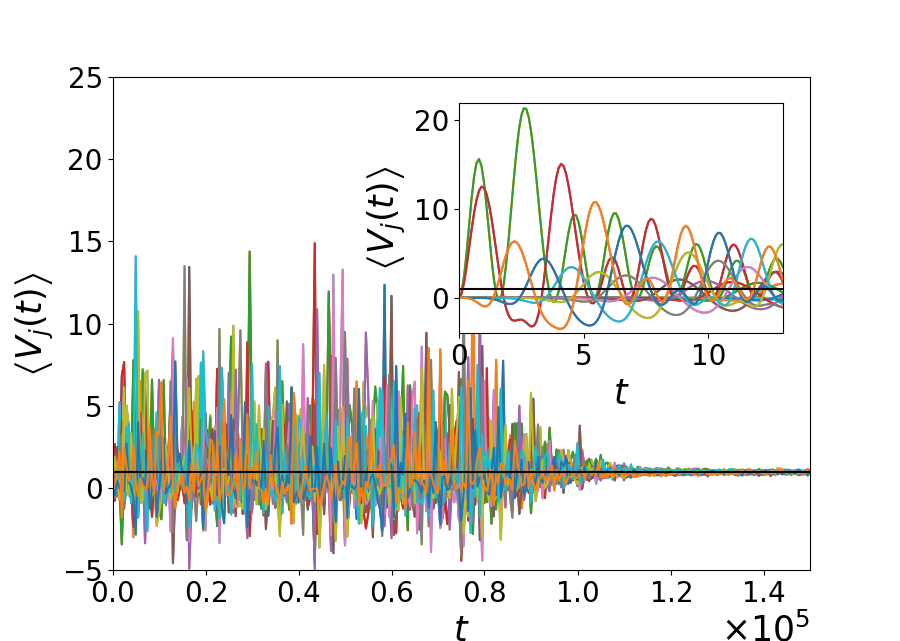}
	\put (-150,183) {$(b)$}
	\hspace{-10mm}
	\caption{$\alpha-$FPUT chain: Same as Fig.~\eqref{kvgp9} but with $\gamma=10^{-8}$. Note that the range of the time $t$ and the averages $\langle T_j \rangle$ and $\langle V_j \rangle$ are different from those in Fig.~\eqref{kvgp9}.}
	\label{kvgpm8}
\end{figure}

\subsection{Evolution of local observables from space localized initial conditions in FPUT chain}

\begin{figure}[ht]
	\centering
	\hspace{-10mm}
	\includegraphics[width=0.55\textwidth]{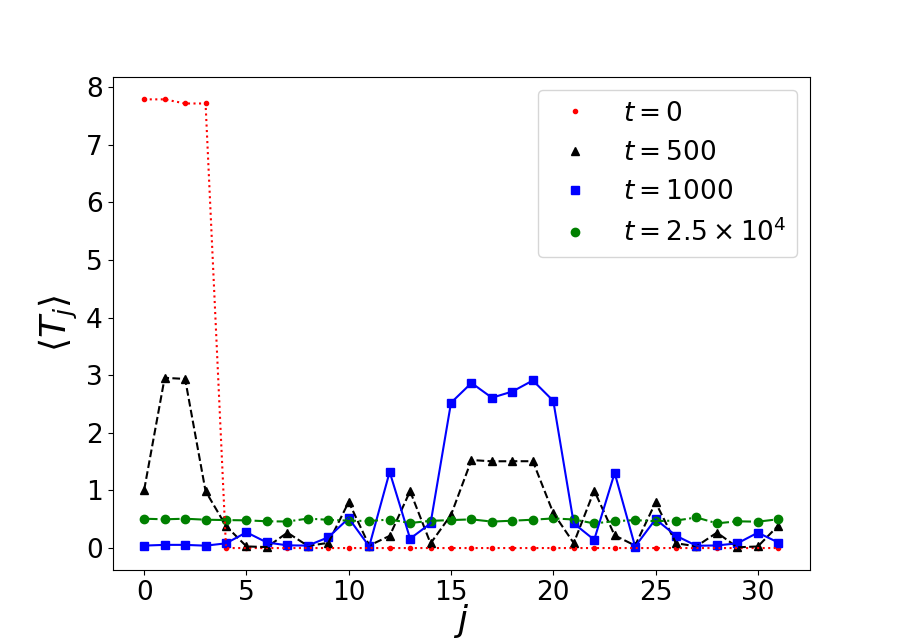}
	\put (-150,183) {$(a)$}
	\hspace{-10mm}
	\includegraphics[width=0.55\textwidth]{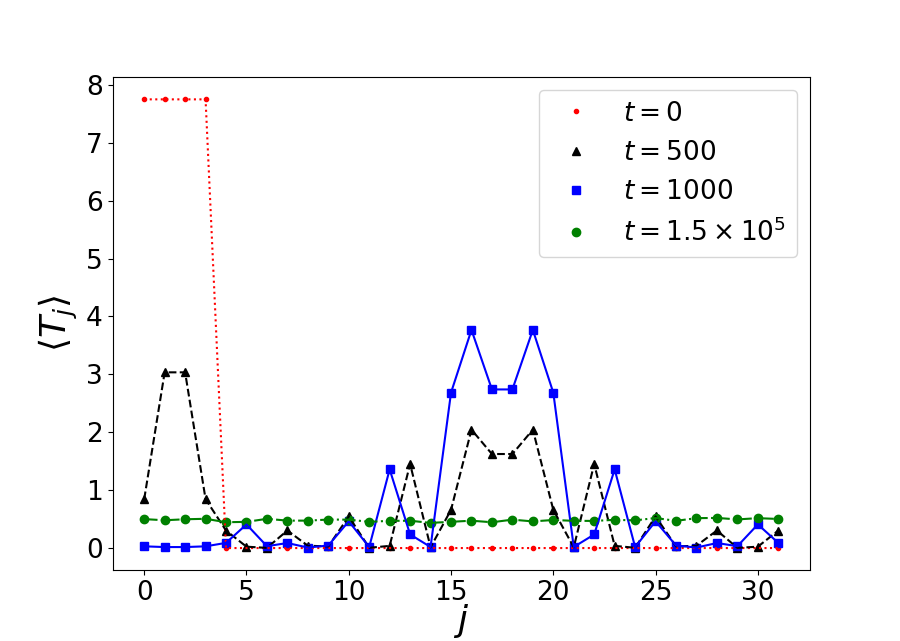}
	\put (-150,183) {$(b)$}
	\hspace{-10mm}
	\caption{$\alpha-$FPUT chain: Spatial profile of $\left\langle T_j\right\rangle$ at different times for   $\gamma=0.9$ (left panel) and $ \gamma=10^{-8}$  (right panel) and other parameters $N = 32$, $E = 31$, $\alpha=0.0848$ and $R=1000$. We see that the initially localized energy quickly spreads through the chain while equipartition is achieved at much longer time scales.}
	\label{evolution}
\end{figure}
The space localized excitations (SLE) are parameterized by the variable $\gamma$ whose magnitude gives an estimate of the width of the initial distribution in phase space. We note that $\gamma=0$ corresponds to a fixed initial state while $\gamma=1$ corresponds to the broadest distribution. In Fig.~(\ref{kvgp9}) we show the time evolution of $\left\langle T_j\right\rangle$ and $\left\langle V_j\right\rangle$ for $\gamma =0.9$.  We see that there is a long transient period and then we see equipartition at times  $\sim 2\times10^4$. At the earliest times, the inset in Fig.~\eqref{kvgp9} shows near-recurrent behaviour. The results for $\gamma =10^{-8}$ are plotted in Fig.~\eqref{kvgpm8}, where we now see that equipartition is achieved at somewhat longer times, around $t \sim 10^5$. In Fig.~\eqref{evolution} we plot the averaged kinetic energy profile at different times. It is seen that the energy spreads quickly through the entire system, while equipartition is achieved at much longer time scales.

\begin{figure}[ht]
	\centering
	\hspace{-10mm}
	\includegraphics[width=0.55\textwidth]{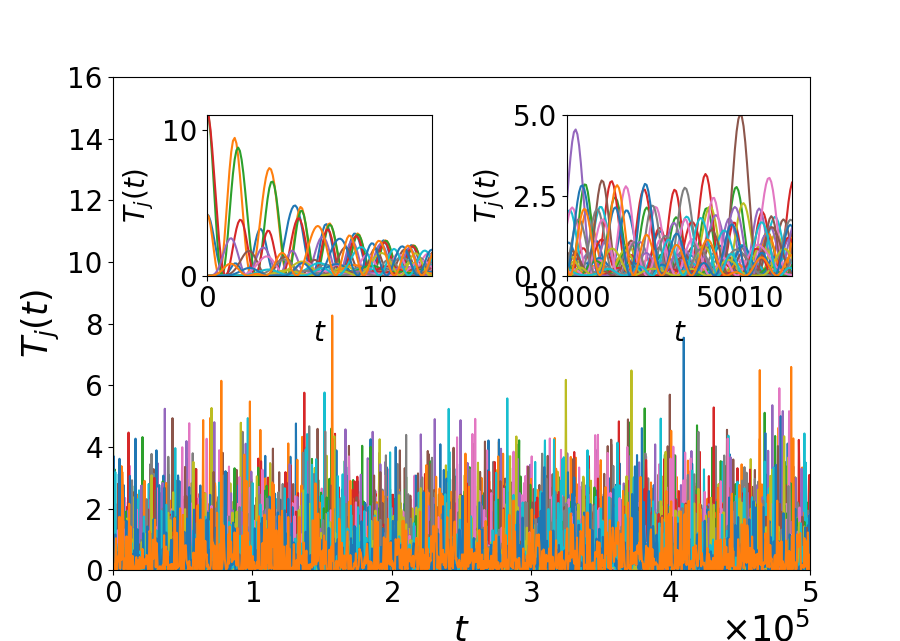}
	\put (-150,183) {$(a)$}
	\hspace{-10mm}
	\includegraphics[width=0.55\textwidth]{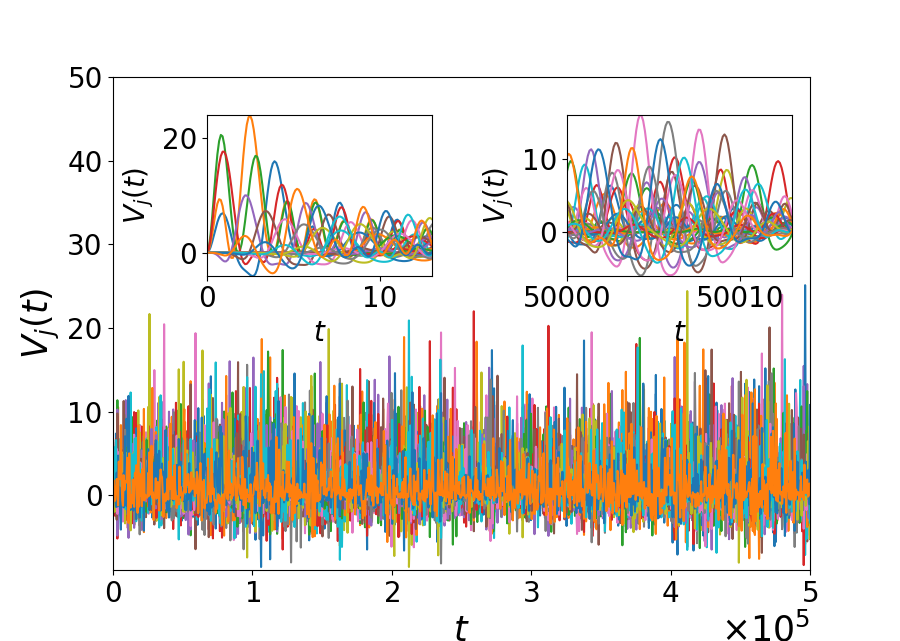}
	\put (-150,183) {$(b)$}
	\hspace{-10mm}
	\caption{$\alpha-$FPUT chain: Plot shows time evolution of $ T_j$ (left panel) and $ V_j$ (right panel) starting from a single initial condition.  Parameter values for this plot are $N=32$, $E=31$, $\alpha= 0.0848$ ($\epsilon=0.0834$). The insets show zoom-ins of the time evolution at early times and at late times and we see oscillatory behaviour in both cases. Thus in this case, no equipartition is achieved.}
	\label{singleR}
\end{figure}

To demonstrate that the averaging procedure is crucial to the equilibration process in this set-up, we show the evolution of  $ T_j$ and $V_j$ for a single initial condition. In this case, we see in Fig.~\eqref{singleR} that no equilibration is achieved and the oscillatory behaviour persists up to $t =5 \times 10^5$. 
\begin{figure}[ht]
	\centering
	\hspace{-10mm}
	\includegraphics[width=0.56\textwidth]{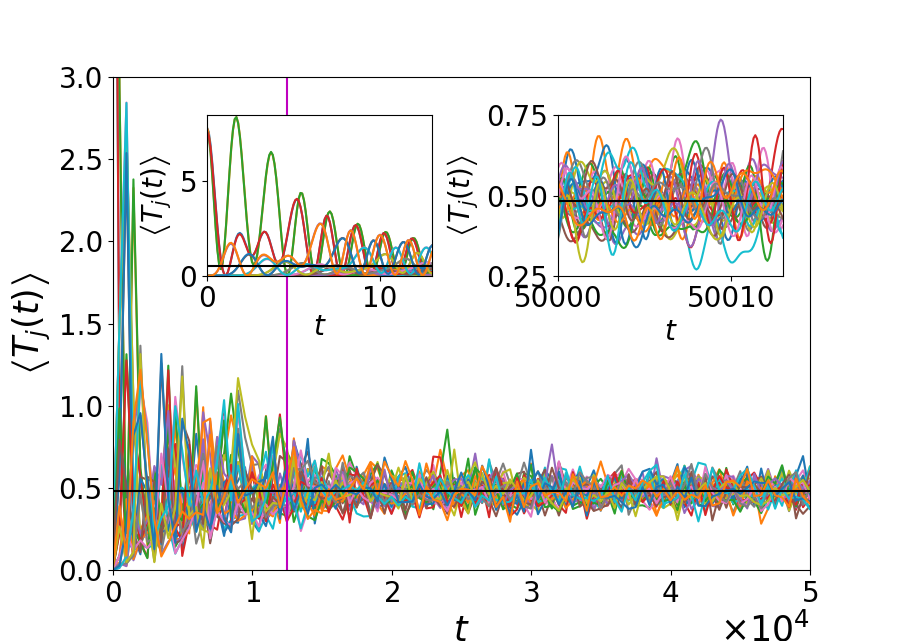}
	\put (-150,186) {$(a)$}
	\hspace{-11.5mm}
	\includegraphics[width=0.443\textwidth]{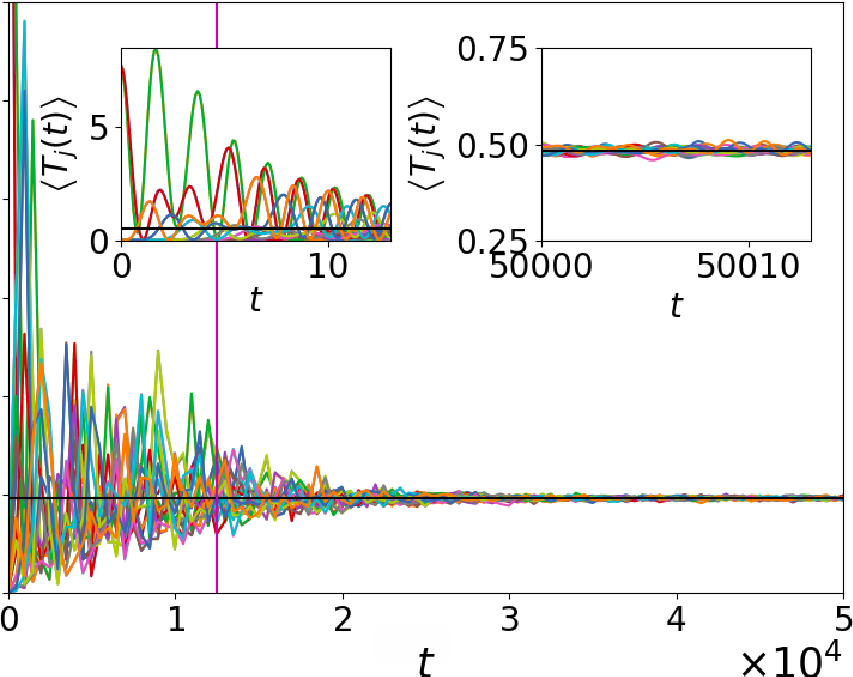}
	\put (-120,186) {$(b)$}
	\hspace{-10mm}
	\caption{$\alpha-$FPUT chain: Here we examine how the fluctuations seen in $\left\langle T_j\right\rangle$ depend on the number of realizations $R$. We plot $\left\langle T_j\right\rangle$  for $R=10^2$ (left panel) and $R=10^4$ (right panel). Other parameters were taken as  $N = 32$, $E = 31$, $\alpha= 0.0848$,  $\gamma=0.9$. The insets show zoom-ins at short and long times. A vertical line is drawn at $t=1.2\times10^4$, which is the equilibration time. Up to this time, both the plots look nearly the same.}
	\label{Rdependence}
\end{figure}

An examination of the plots in Figs.~(\ref{kvgp9},\ref{kvgpm8}) shows that even at late times, the averaged quantities continue to fluctuate around their equilibrium values. We show 
in Fig.~\eqref{Rdependence} that these fluctuations in fact decrease with increase in the number of realizations $R$ used to compute averages. We also see that  $\left\langle T_j\right\rangle$ at pre-thermalization times does not depend significantly on $R$ and the plots are nearly identical for $R=10^2$ and $R=10^4$ (up to the vertical line in Fig.~\eqref{Rdependence}).

\begin{figure}[htp]
	\centering
	\hspace{-10mm}
	\includegraphics[width=0.55\textwidth]{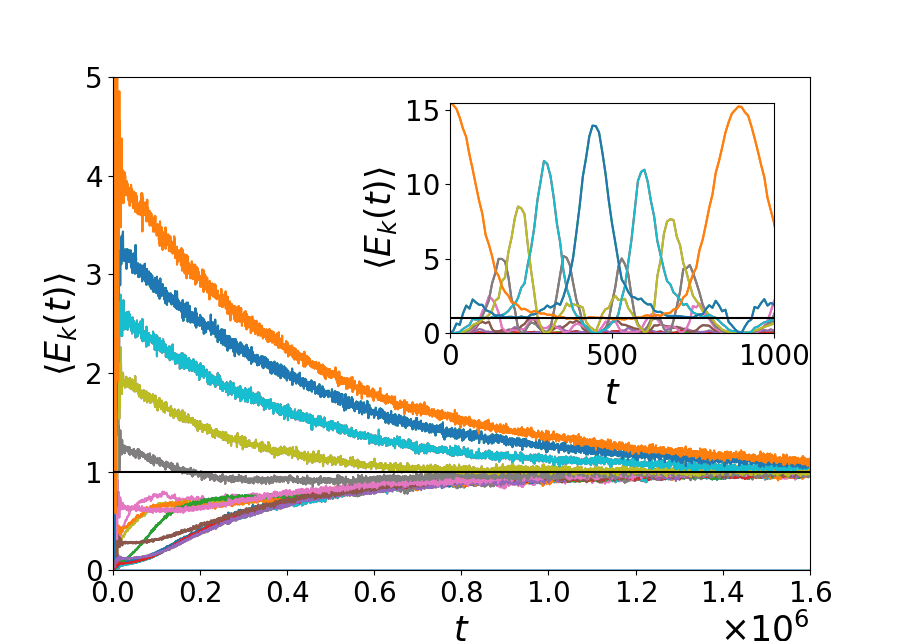}
	\put (-150,183) {$(a)$}
	\hspace{-10mm}
	\includegraphics[width=0.55\textwidth]{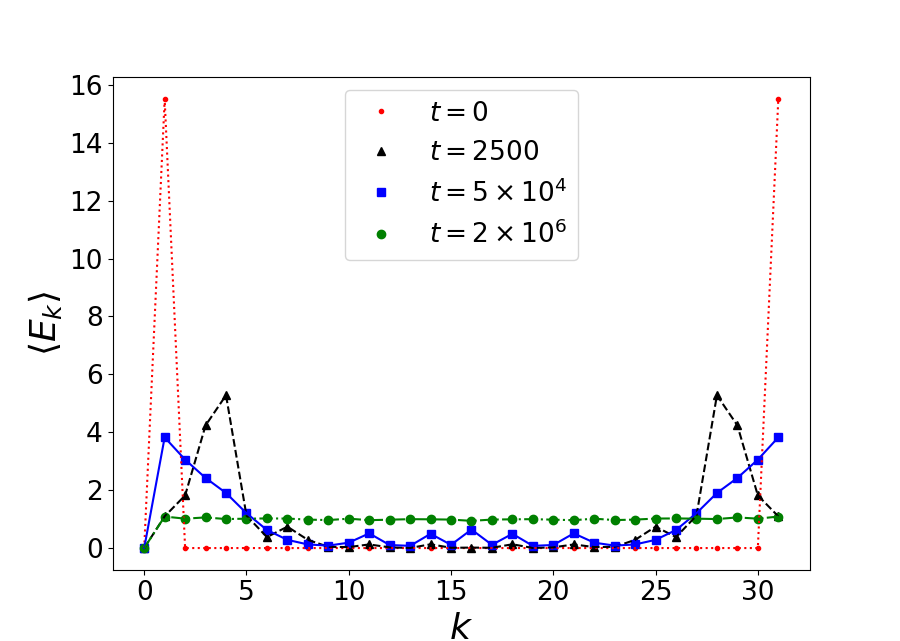}
	\put (-150,183) {$(b)$}
	\hspace{-10mm}
	\caption{$\alpha-$FPUT chain: Left panel shows time evolution of $\left\langle E_k\right\rangle$  for $N=32, E=31, \alpha=0.0848$ and  $R=1000$. Only modes $k=1,31$ are initially excited. The inset shows quasiperiodic behaviour at short time scale. Right panel shows the mode energy profile at different  times.}
	\label{NMevol}
\end{figure}
\begin{figure}[ht]
	\centering
	\includegraphics[width=0.55\textwidth]{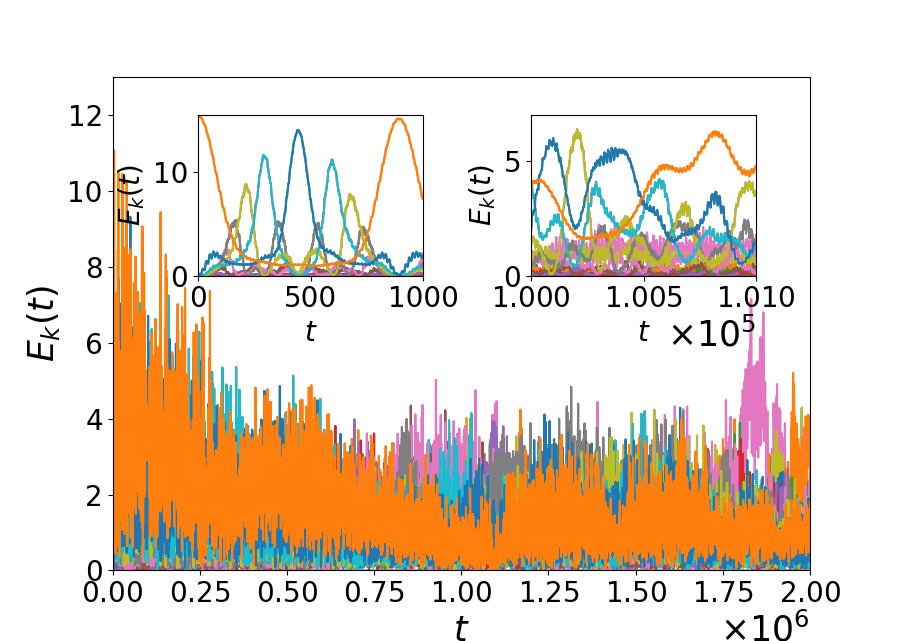}
	\caption{$\alpha-$FPUT chain: Plot shows time evolution of $E_k $  starting from a single initial condition with energy in two normal modes.  Parameter values for this plot are $N=32$, $E=31$, $\alpha= 0.0848$. The insets show zoom-ins  of  the time evolution at early times and at late times and we see oscillatory behaviour in both cases. Thus in this case, no equipartition is achieved.}
	\label{NMsingle}
\end{figure} 

\subsection{Normal mode localized initial conditions}
We now discuss and compare our results with those obtained in studies on equipartition of normal mode energies using initial conditions which were investigated in \cite{Onorato2015}. 
The authors in \cite{Onorato2015}  considered  initial conditions  where the  energy was  distributed between the modes $k=1$ and $k=31$ with frequencies  $\Omega_{1}=\Omega_{31}$.  Averages were done over $1000$ initial conditions by choosing  random phases for 
the modes.  The time evolution of the normal mode energies was monitored to check for equipartition. Here we reproduce their numerical results and compare with the results in the previous section. We consider again  $N=32$ particles with total energy $E=31$. In Fig.~\eqref{NMevol} we show the time evolution of the energy of all the modes in the system. Comparing with Figs.~(\ref{kvgp9},\ref{kvgpm8}) it is clear that equilibration now occurs at a  time scale ($\sim 2 \times 10^6$) that is about an order of magnitude  longer. In Fig.~(\ref{NMsingle}) we see again that one does not see any signs of equilibration in the evolution of a single realization.

\subsection{Comparison with temporal averaging protocol}
\begin{figure}[ht]
	\centering
	\hspace{-10mm}
	\includegraphics[width=0.55\textwidth]{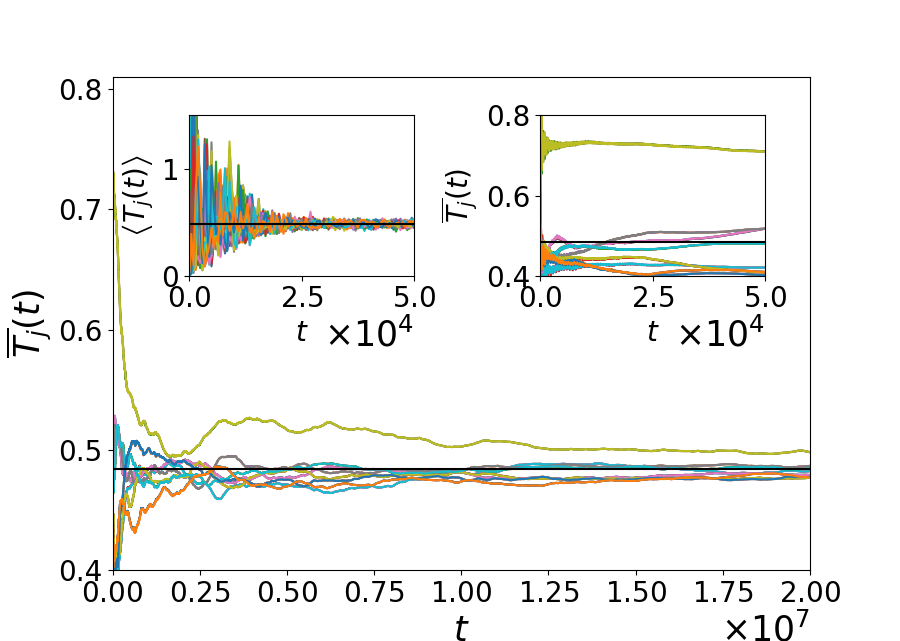}
	\put (-150,183) {$(a)$}
	\hspace{-9.4mm}
	\includegraphics[width=0.55\textwidth]{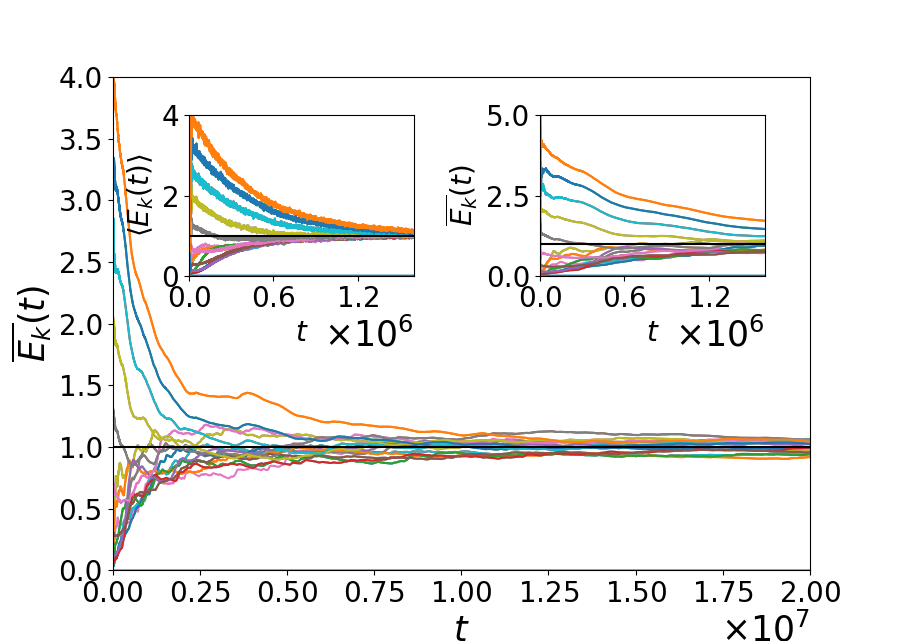}
	\put (-150,183) {$(b)$}
	\hspace{-10mm}
	\caption{$\alpha-$FPUT chain: Plot shows time evolution of running time averages of $ T_j$ for SLE (left panel) and $ E_k $ for NMLE (right panel) starting from a single initial condition.  Parameter values for this plot are $N=32$, $E=31$, $\alpha= 0.0848$ ($\epsilon=0.0834$).  The insets show comparison between time averages and ensemble averages and illustrates that the latter procedure leads to faster thermalization.}
	\label{Timeaverage}
\end{figure}

As discussed in Sec.~\ref{ssec:avg-proc} one can discuss thermalization using a different protocol where one starts with a single initial condition and then considers a time average of any given observable. This is given by Eq.~\eqref{eq:tavg}. In Fig.~\eqref{Timeaverage} we show results obtained using this protocol for both the space-local and normal mode observables. The insets in the figures show that thermalization time scales are completely different from those obtained by the ensemble averaging protocol.

\subsection{Estimation of equilibration time from  entropy}

\begin{figure}[ht]
	\centering
	\hspace{-10mm}
	\includegraphics[width=0.55\textwidth]{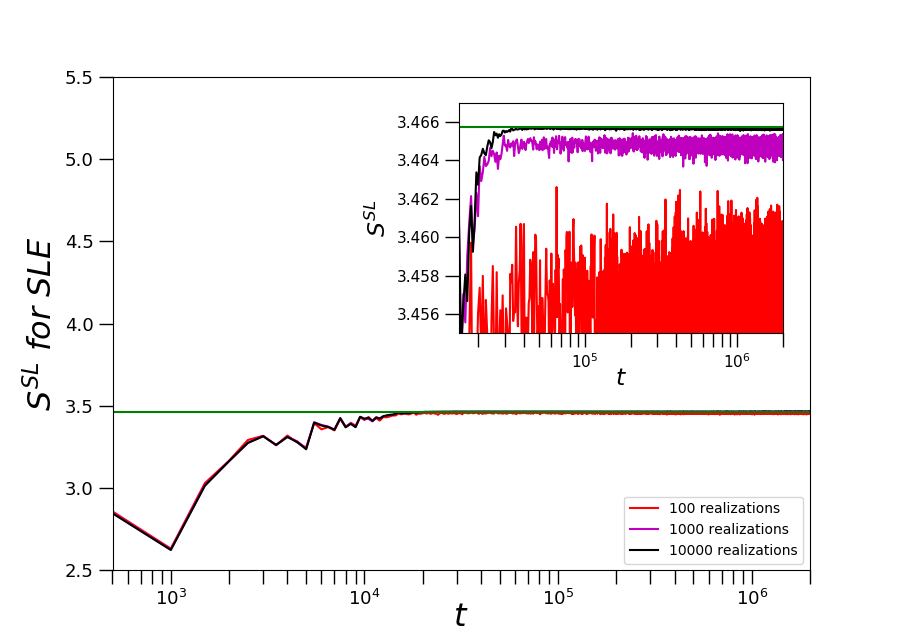}
	\put (-150,183) {$(a)$}
	\hspace{-10mm}
	\includegraphics[width=0.55\textwidth]{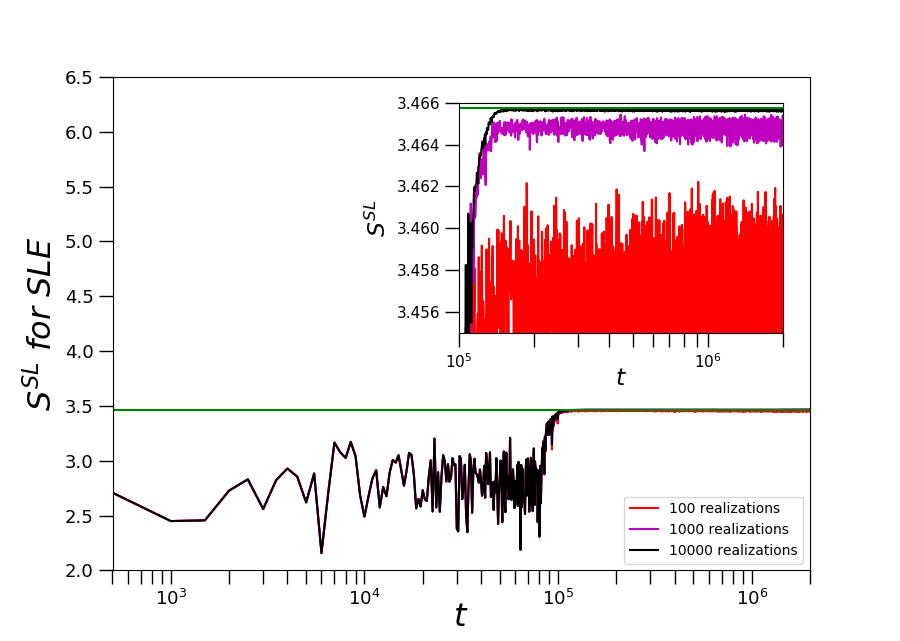}
	\put (-150,183) {$(b)$}
	\hspace{-10mm}
	\caption{$\alpha-$FPUT chain: Plot of $S^{SL}(t)$ corresponding to $\left\langle T_j\right\rangle$ for $\gamma=0.9$ (left panel) and $\gamma=10^{-8}$ (right panel) with other parameters $N=32, E=31, \alpha=0.0848$. The insets show the convergence of the equilibration process as the number of realizations is increased.  }
	\label{EntropyEv}
\end{figure}

To get a more systematic and quantitative estimate of the equilibration time we now look at the entropy function defined in Eq.~\eqref{eq:ent}. This in some sense, performs an average over all the degrees of freedom and attains its maximum value $\ln N$ when all degrees have equilibrated. 
In Fig.~\eqref{EntropyEv}, we plot the evolution of entropy for the two parameter values $\gamma=0.9$ and $\gamma=10^{-8}$. The insets show zoom-ins near the equilibrium value $\ln N$, showing the approach to  equilibration and its dependence on the number of realizations. For higher number of realizations, the fluctuations in the entropy is lower and also the mean is closer to the equilibrium value. 

\begin{figure}[ht]
	\centering
	\includegraphics[width=0.6\textwidth]{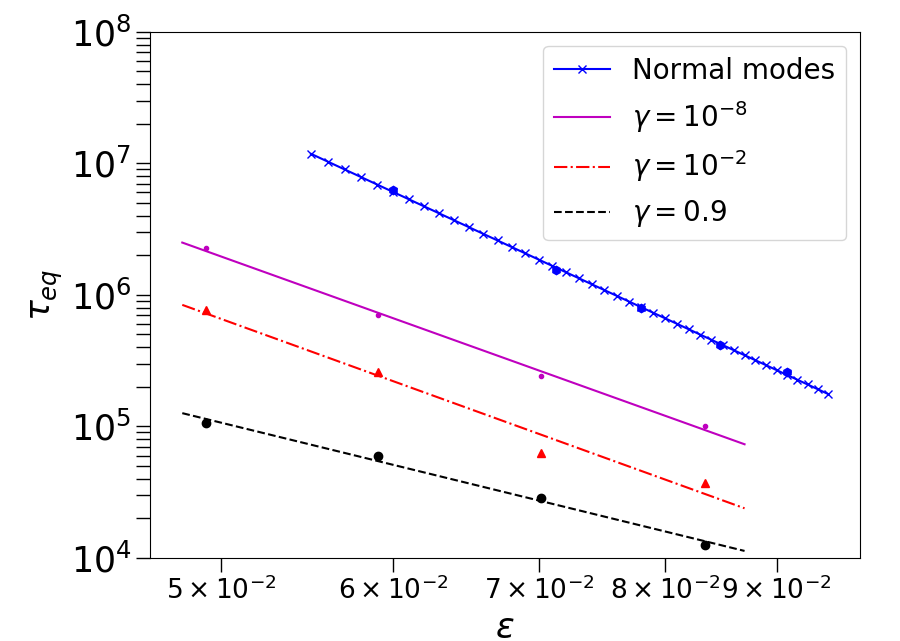}
	\caption{$\alpha-$FPUT chain: Graph showing $\tau_{\rm eq}$ of $\left\langle T_j\right\rangle$  for $N=32,E = 31$ as a function of $\epsilon$. The slopes of the fitting lines give  $a=5.9, 6.0, 4.0$  for $\gamma=10^{-8}, 0.01$ and $0.9$ respectively. We also show the equilibration times obtained from the normal mode entropy function $S^{NML}$ for normal mode localized initial conditions, which leads to an exponent $a\approx 7.7$.}
	\label{relaxationtimeepsilon}
\end{figure}

We  use the criterion of Eq.~\eqref{criterion} to estimate the equilibration time from the entropy $S^{SL}$ corresponding to $\langle T_j\rangle$ and find $\tau_{\rm eq}\approx100300$ and $12500$ for $\gamma=10^{-8}$ and $0.9$ respectively.
In general we find that as the "width" of the distribution $\gamma$ is increased, thermalization is found to happen faster. We will discuss this again in Sec.~\ref{sec:chaos}. We  compute the equilibration time for different values of the dimensionless parameter $\epsilon$ for $N=32$. These results are plotted in Fig.~\eqref{relaxationtimeepsilon} where we find a power-law dependence of equilibration time, $\tau_{\rm eq}$, on $\epsilon$  of the form
\begin{equation}\label{tauform}
\tau_{\rm eq} \propto \frac{1}{\epsilon^{a}}~,
\end{equation}
The value $a$ is found to depend on $\gamma$ and lies between 4 and 6. This is significantly different from the form 1/$\epsilon^8$ obtained in \cite{Onorato2015}, by considering equilibration of normal modes. In Fig.~\eqref{relaxationtimeepsilon} we also indicate the relaxation time results for the normal modes which give  $a \approx 7.7$. 

\begin{figure}[!]
	\centering
	\includegraphics[width=1.1\textwidth]{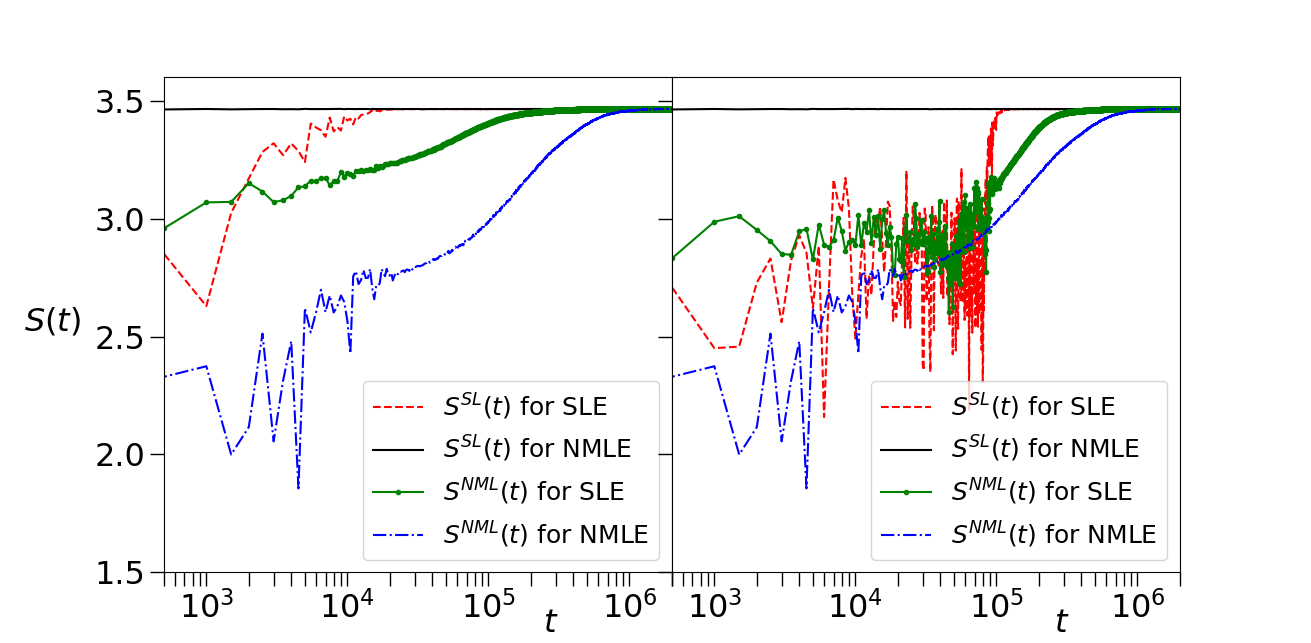}
	\put (-380,250) {$(a)$}
	\put (-163,250) {$(b)$}
	\caption{$\alpha-$FPUT chain: Graphs showing the evolution of the entropy corresponding to space-local observables and normal modes, for $\gamma=0.9$ (left panel) and  $\gamma=10^{-8}$ (right panel), with other parameters given by  $N=32, E=31, \alpha=0.0848$.  Results for both space localized initial conditions (SLE) and normal mode localized initial conditions (NMLE) are shown.}
	\label{combo8}
\end{figure}

It is to be expected that the equilibration time scale should depend  not only on the initial ensemble in which the system is prepared, but also on the observable for which equipartition is being tested. We investigate this question further by computing the entropy functions $S^{SL}$ and $S^{NML}$ for both types of initial conditions, namely space localized (SLE) and normal mode localized (NMLE). 
These results have been plotted in Fig.~\eqref{combo8}. We see clearly that the relaxation of normal mode coordinates is slower than that of the space localized observables, irrespective of initial distribution.


\subsection{Comparison with harmonic chain and Toda chain}
\label{ssec:harmonic-toda}
\begin{figure}[ht]
	\centering
	\hspace{-10mm}
	\includegraphics[width=0.55\textwidth]{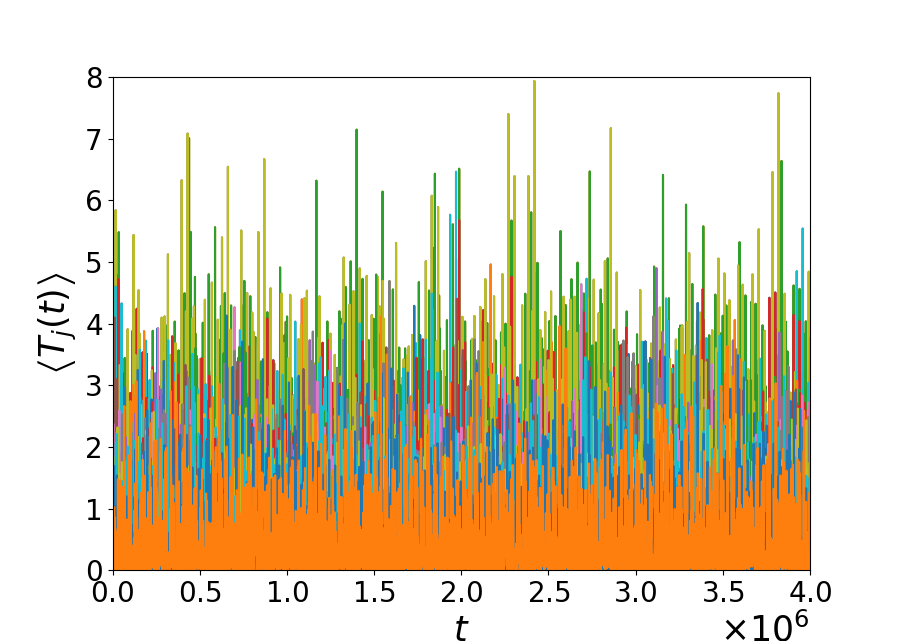}
	\put (-150,183) {$(a)$}
	\hspace{-10mm}
	\includegraphics[width=0.55\textwidth]{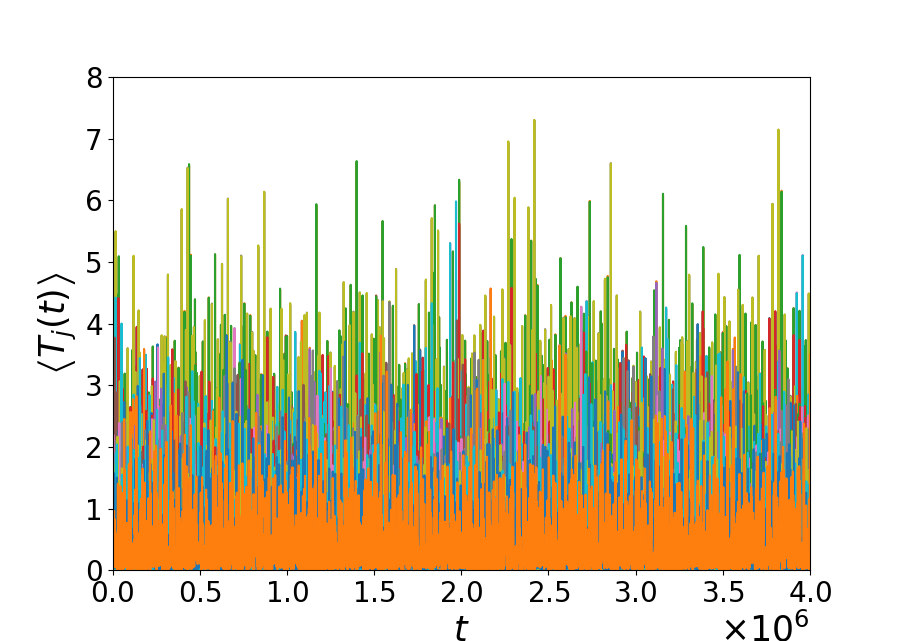}
	\put (-150,183) {$(b)$}
	\hspace{-10mm}
	\caption{Harmonic chain: Plot of $\left\langle T_j\right\rangle$ as a function of time, for $N = 32, E = 31, R=1000$ and $\gamma=10^{-8}$ (left panel) and $\gamma=0.9$ (right panel). In both cases there is no sign of thermalization.}
	\label{KEharmonic}
\end{figure}

\begin{figure}[ht]
	\centering
	\hspace{-10mm}
	\includegraphics[width=0.55\textwidth]{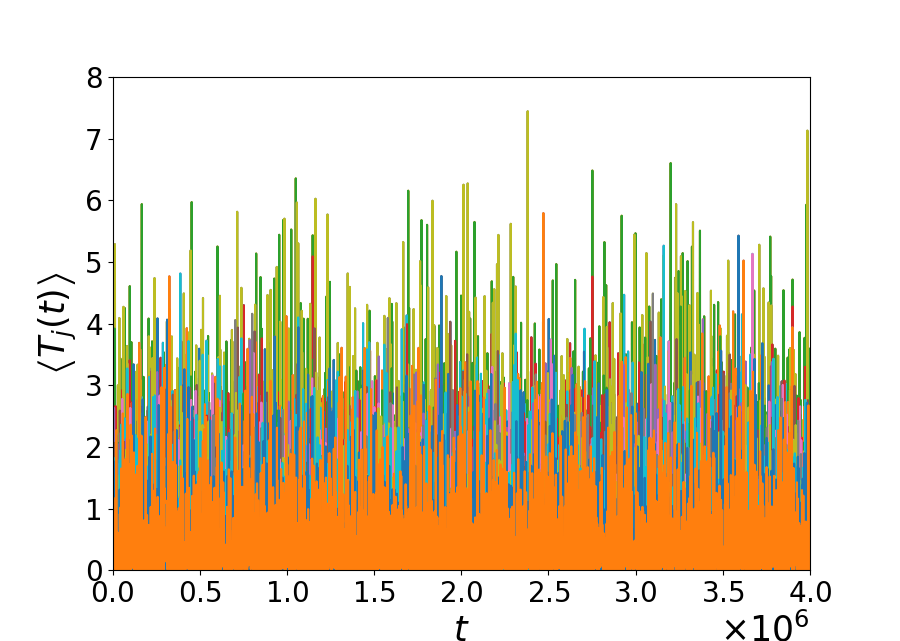}
	\put (-150,183) {$(a)$}
	\hspace{-10mm}
	\includegraphics[width=0.55\textwidth]{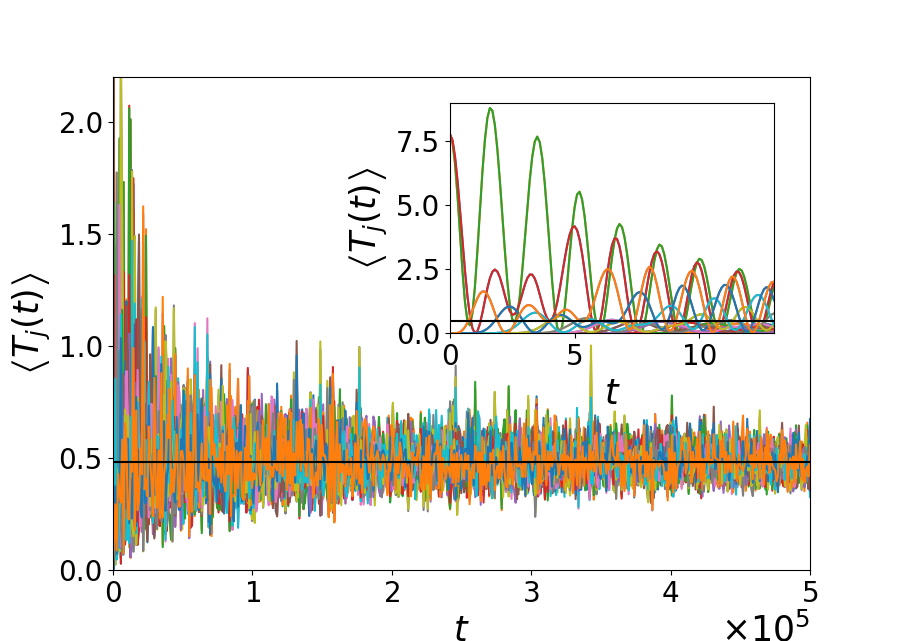}
	\put (-150,183) {$(b)$}
	\hspace{-10mm}
	\caption{Toda chain: Plot of $\left\langle T_j\right\rangle$ as a function of time, for $N = 32, E = 31, \alpha=0.0848, b=2\alpha, g=b^{-1}, R=1000$ and $\gamma=10^{-8}$ (left panel) and $\gamma=0.9$ (right panel). Now we observe thermalization in the right panel.}
	\label{KEtoda}
\end{figure}

To investigate the role of integrability, we now repeat the above computations in two integrable models that are related to the FPUT system in the limit of weak nonlinearity. We consider the harmonic chain which is described by the Hamiltonian in Eq.~\eqref{eq:ham}, with $\alpha=0$, and the Toda chain, described by the Hamiltonian
\begin{equation}\label{eq:todaham}
H(\boldsymbol{p},\boldsymbol{q})=\sum_{i=0}^{N-1}\left[\frac{p_i^2}{2}+ \frac{g}{b} e^{b (q_{i+1}-q_{i})} \right]~. 
\end{equation}
The Toda system is known to be integrable \cite{Toda1967,Toda1975} and has been much studied as the integrable limit of the FPUT chain \cite{Benettin2013,Casetti1996,Goldfriend2019,Fu2019}. 
The parameter choice $b=2\alpha$ and $g=b^{-1}$ would then approximate the $\alpha-$FPUT potential to leading nonlinearity.  Starting with the   same space-localized initial conditions as in the previous sections, we now check equipartition of kinetic energy $T_i$. In Fig.~\eqref{KEharmonic} we see that no equilibration is achieved for the harmonic chain. On the other hand, somewhat surprisingly, we see in Fig.~\eqref{KEtoda} that the Toda chain does equilibrate, provided we start with a wider initial distribution ($\gamma=0.9$).


\begin{figure}[ht]
	\centering
	\hspace{-15mm}
	\includegraphics[width=0.53\textwidth]{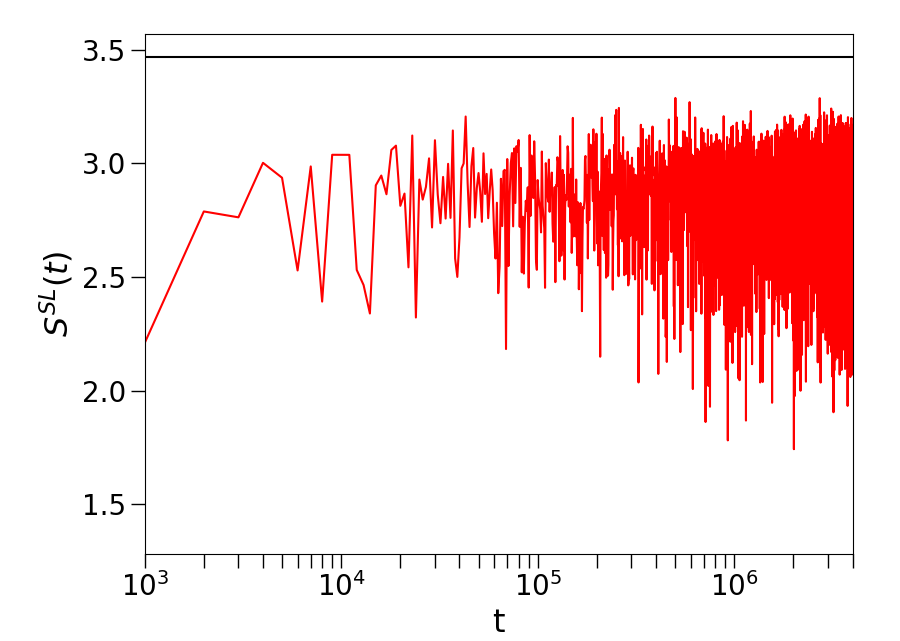}
	\put (-140,190) {$(a)$}
	\hspace{-8mm}	
	\includegraphics[width=0.53\textwidth]{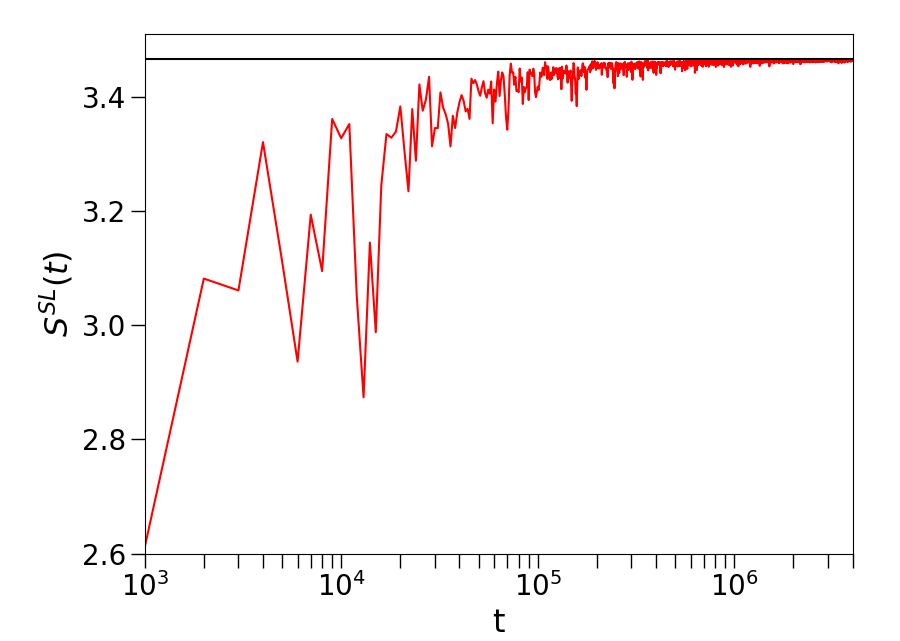}
	\put (-140,190) {$(b)$}	
	\hspace{-15mm}
	\caption{Toda chain: Plot of the entropy $S^{SL}(t)$ for $\left\langle T_j\right\rangle$ of the Toda chain with same parameters as Fig.~\ref{KEtoda} and $\gamma=10^{-8}$ (left panel) and $\gamma=0.9$ (right panel). The second case shows clear equilibration.}
	\label{Enttoda}
\end{figure}

In Fig.~\eqref{Enttoda} we plot the entropy function $S^{SL}$ and using the criterion in Eq.~\eqref{criterion}, estimate the equilibration time and find $\tau_{\rm eq} \approx 66000$ for $\gamma=0.9$. The dependence of the equilibration time ($\tau_{\rm eq}$) on $\gamma$ will be discussed quantitatively in the next section.

\section{Relation to chaos}
\label{sec:chaos}
We will now argue that  the process of equilibration in the FPUT  protocol is intimately connected to the  growth of perturbations of initial condition, and hence to chaos. This idea then also explains the absence of equilibration in the harmonic chain and the slow equilibration in the Toda chain.  
\begin{figure}[ht]
	\centering
	\includegraphics[width=\textwidth]{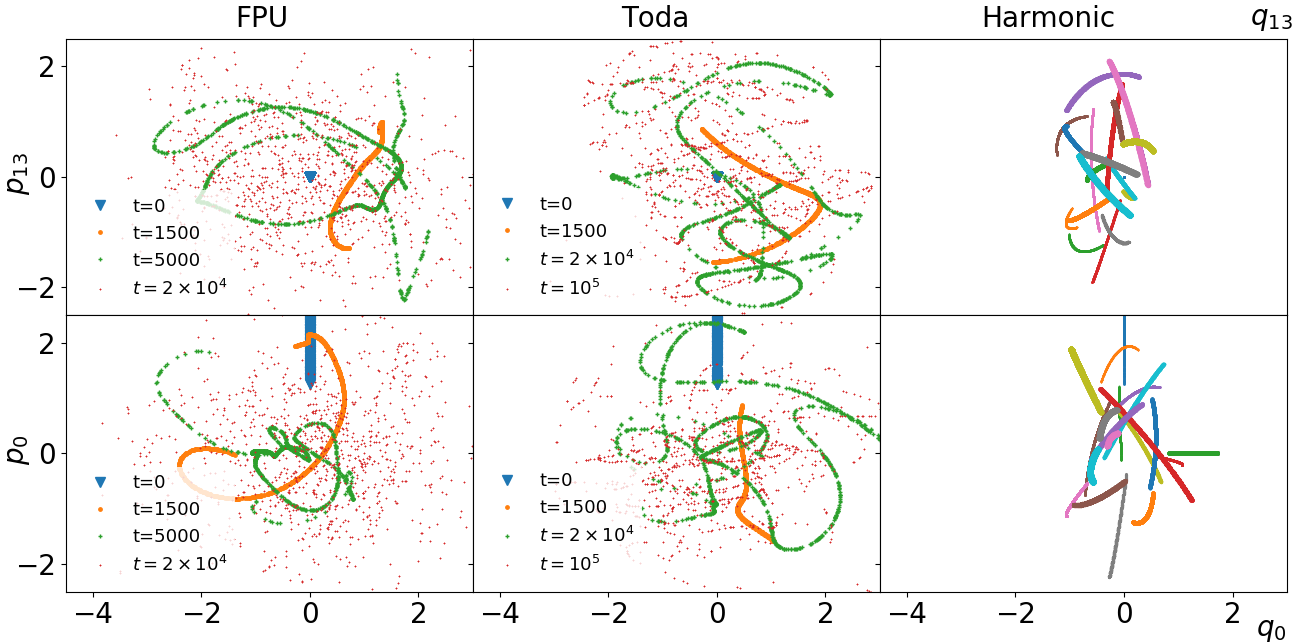}
	\caption{The time-evolution of the ensemble of initial conditions obtained using Eqs.\eqref{eq:ic1}-\eqref{eq:ic2} is shown for the FPUT chain with $\alpha = 0.0848$ (left column), the corresponding Toda chain (middle column), and the harmonic chain (right column), in the $(p_{13},q_{13})$ and $(p_0,q_0)$ planes (top and bottom rows, respectively). The different lines for harmonic chain are at increased times, with increasing marker size.}
	\label{schematic}
\end{figure}
We first quantify the growth of perturbations in the system. Let us consider an infinitesimal perturbation $\{ \delta q_i(0),\delta p_i(0) \}$ of an initial condition $\{ q_i(0), p_i(0)\}$. We compute the quantity
\begin{align}
Z(t)&=\frac{\sum_{i=1}^N \left[ \delta q_i^2(t) +\delta p_i^2(t) \right]}{\sum_{i=1}^N \left[ \delta q_i^2(0) +\delta p_i^2(0) \right]}\,.
\end{align}
We then compute the ensemble averaged time-dependent Lyapunov exponent $\lambda(t)$ defined as
\begin{align}
\lambda(t) =\frac{1}{2t} \langle \ln Z(t) \rangle~,
\end{align}
where  $\langle...\rangle$ denotes an average over initial conditions  $\{ q_i(0), p_i(0)\}$ chosen from the distribution $\rho_0({\bf q},{\bf p})$. As described earlier, the numerical integration of $\{ q_i(t), p_i(t), \delta q_i(t),\delta p_i(t) \}$ is done by solving $2N + 2N$ nonlinear and linearized equations. The largest Lyapunov exponent $\Lambda$ is then given by $\Lambda = \lim_{t \to \infty} \lambda(t)$. For a harmonic chain,  transforming to normal modes shows that $Z(t)$ is bounded. For a Toda chain, a transformation to action angle variables suggests that $Z(t)$ should grow linearly with time. This is consistent with $\Lambda=0$ for both the harmonic chain and the Toda chain.

The basic picture that illustrates the difference between the three models is shown in Fig.~\eqref{schematic}, where we show the time-evolution of the ensemble of initial conditions obtained using Eqs.\eqref{eq:ic1}-\eqref{eq:ic2}. The FPUT chain shows a fast growth in phase space because of its positive Lyapunov exponent, while it takes much longer for the Toda chain because of its integrability (and linear temporal growth of perturbations). There is no spread in the harmonic chain. 

\begin{figure}[ht]
	\centering
	\hspace{-10mm}
	\includegraphics[width=0.55\textwidth]{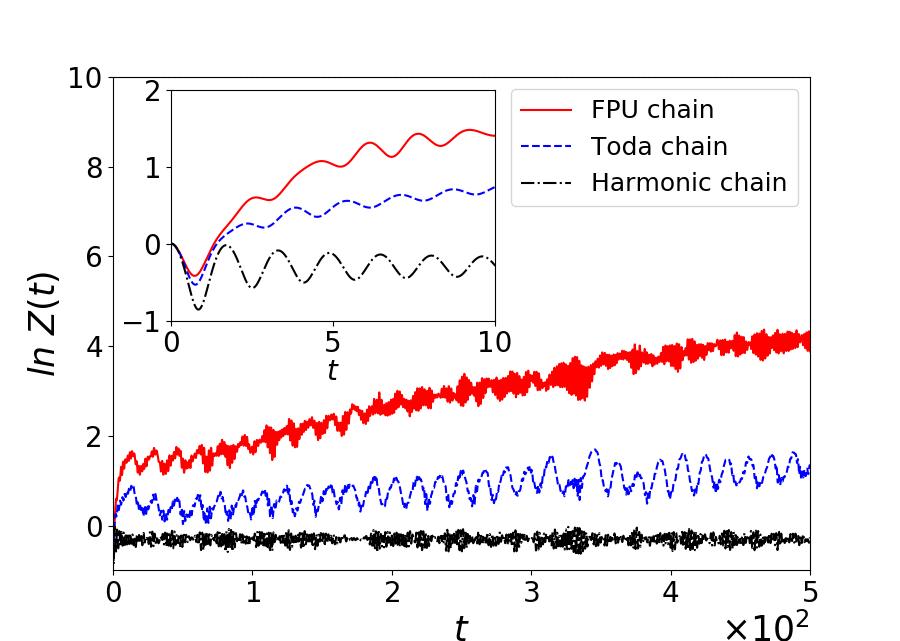}
	\put (-150,183) {$(a)$}
	\hspace{-10mm}
	\includegraphics[width=0.55\textwidth]{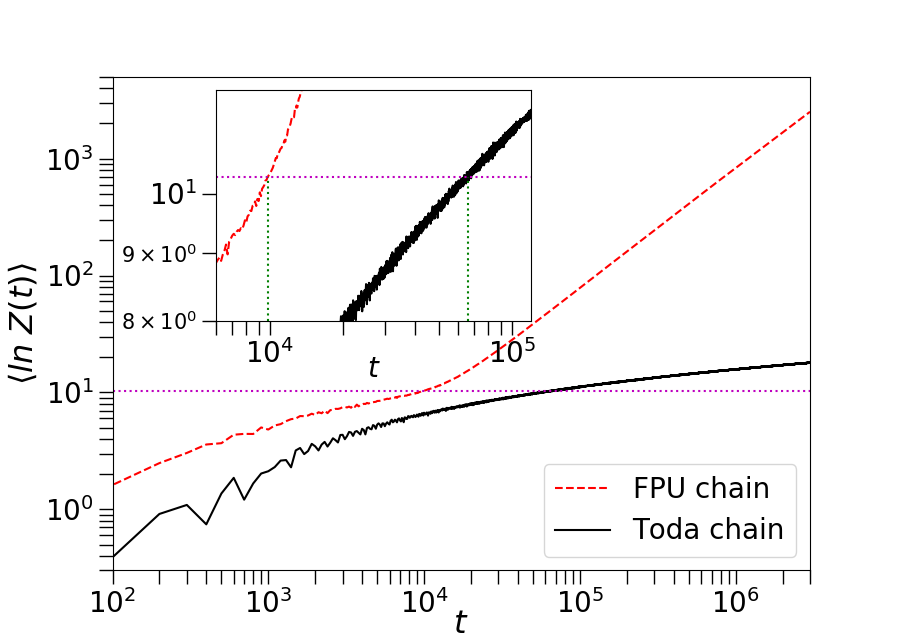}
	\put (-150,183) {$(b)$}
	\hspace{-10mm}
	\caption{Left panel shows $Z(t)$ of $\alpha-$FPUT chain, Toda chain and the harmonic chain at short times. It can be seen that the growth of perturbations is bounded for the harmonic chain and linear for the Toda chain. Right panel shows $Z(t)$ of $\alpha-$FPUT chain and the Toda chain for long times. The growth of perturbations in the $\alpha-$FPUT chain though linear at short times, becomes exponential at long times. It can be seen that the growth of perturbations is linear for the FPU chain and logarithmic for the Toda chain in the logarithmic scale. A horizontal line drawn at $\langle \log Z(t) \rangle=10.3$ intersects both the curves. Its significance is explained in the text.}
	\label{lyap-comp}
\end{figure}

In Fig.~\eqref{lyap-comp} we plot $\langle \log Z(t) \rangle$ for the FPUT chain as well for the corresponding harmonic chain and Toda chain (with $b=2 \alpha$, $g=b^{-1}$). We confirm the expected exponential growth of $Z(t)$ for the FPUT chain at large times, it's linear growth for the Toda chain and the lack of growth in the case of harmonic chain. 
In the inset of the right panel, we also show the line $\langle \log Z(t) \rangle = 10.3$, which corresponds to the equilibration time of the  $\alpha-$FPUT chain. It turns out that the point of intersection of this horizontal line with the Toda chain is very close to its equilibration time. This gives us a means to relate the thermalization properties of both the systems to the growth of their perturbations.

\begin{figure}[ht]
	\centering
	\includegraphics[width=0.6\textwidth]{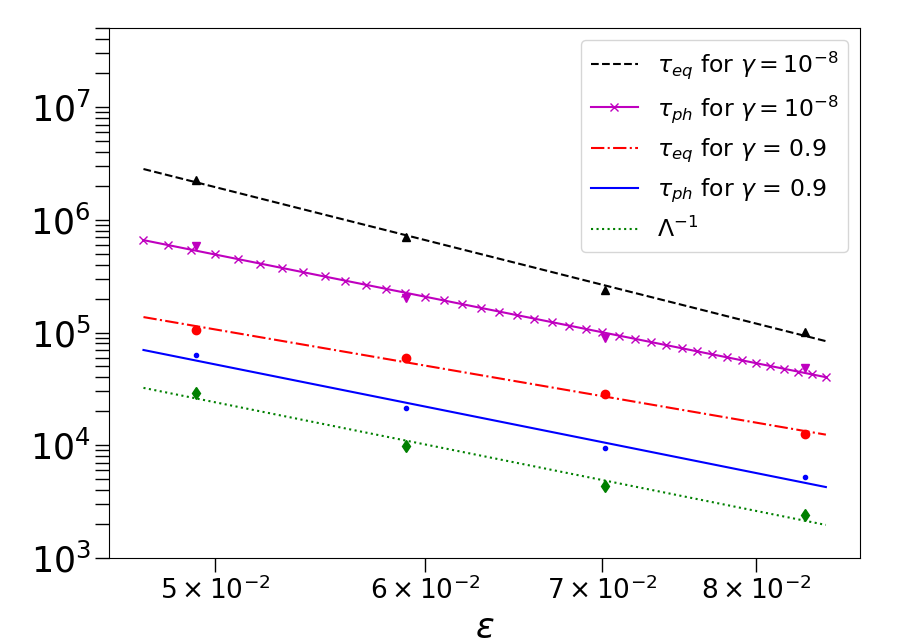}
	
	\caption{$\alpha-$FPUT chain: Graph showing equilibration time $\tau_{\rm eq}$, the phase-space covering time scale $\tau_{ph}$ for $ \gamma=0.9$ and $ \gamma=10^{-8}$ and the inverse $\Lambda^{-1}$ of the maximal Lyapunov exponent of $\left\langle T_j\right\rangle$ as a function of $\alpha$.  Other parameters $N = 32$, $E = 31$  and $R=1000$. The slopes of the five lines (from top to bottom) are $-5.9, -4.7, -4.0, -4.7, -4.7$.} 
	\label{tauvseps}
\end{figure}

In order to explore a possible relation between the equilibration time $\tau_{\rm eq}$ and the maximal Lyapunov exponent $\Lambda$, we show, in Fig.~\eqref{tauvseps}, the dependence of  $\tau_{\rm eq}$ and $\Lambda^{-1}$ on the nonlinearity parameter $\epsilon$ for $\gamma=0.9$ and $\gamma= 10^{-8}$. The slopes are close to each other, suggesting a possible relation between the equilibration time $\tau_{\rm eq}$ and the Lyapunov exponent $\Lambda$. We propose that the close relation between the scaling of  $\tau_{\rm eq}$ and $\Lambda^{-1}$ with respect to $\epsilon$ can be understood through the following argument which relates the growth of perturbations of initial conditions to thermalization.

\begin{figure}[ht]
	\centering
	\hspace{-10mm}
	\includegraphics[width=0.545\textwidth]{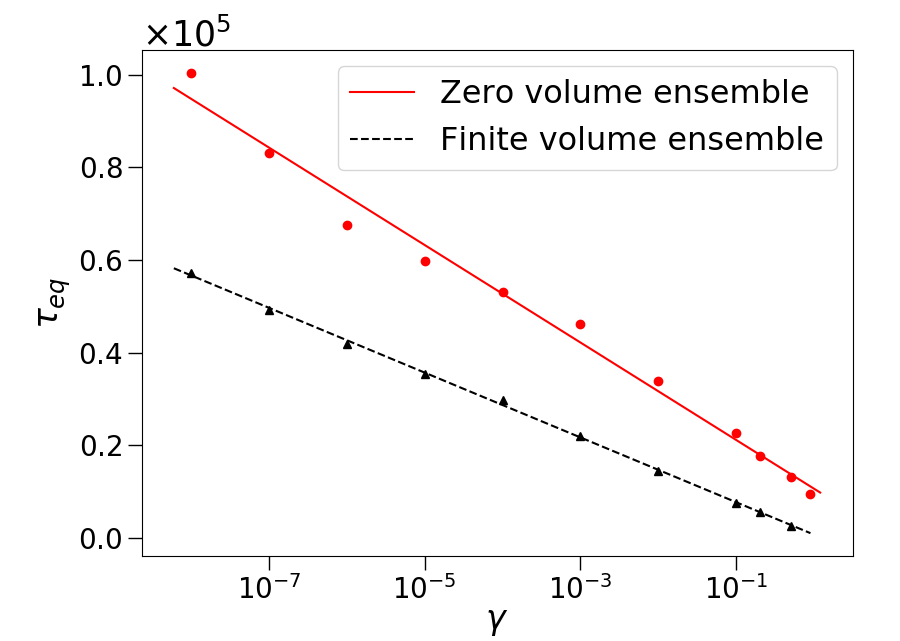}
	\put (-125,188) {$(a)$}
	\hspace{-5mm}
	\includegraphics[width=0.53\textwidth]{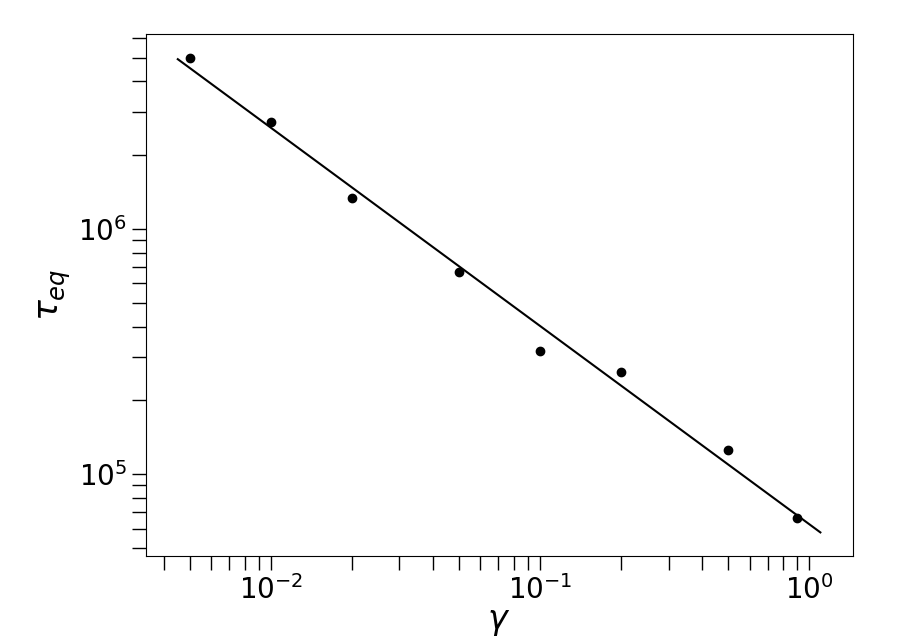}
	\put (-125,188) {$(b)$}
	\hspace{-10mm}
	\caption{Graphs showing the dependence of equilibration time of $\left\langle T_j\right\rangle$ on $\gamma$ for the $\alpha$-FPUT chain (left panel) and the Toda chain for $\alpha=0.0848$ (right panel). Other parameters are $N = 32$, $E = 31$,  and $R=1000$.  For the $\alpha$-FPUT chain, plotted on a linear-log scale, there are two different ensembles. The solid line represents an ensemble of initial conditions occupying zero volume in the phase space. It is described by Eqs.~\eqref{eq:ic1}-\eqref{eq:ic2}. The slope of this line is $-4.5\times 10^3$. The dashed line represents an ensemble of initial conditions occupying a finite volume in the phase space. The slope of this line is $-3.0\times 10^3$. The slope for the Toda  chain, plotted on a log-log scale, is $-0.81$.}
	\label{Fchaos}
\end{figure}

Let us define $D(t)=\left[\sum_{i=1}^N \left( \Delta q_i^2(t) +\Delta p_i^2(t) \right) \right]^{1/2}$ to be the distance at time $t$, of points that are initially separated by a small but {\emph{finite}} separation $D(0)$, due to a finite perturbation of initial conditions. 
In particular, we will interpret $D(t)$ as the ``spread'' of an ensemble of trajectories, for example, using the ensemble of initial conditions as described in Eqs.~\eqref{eq:ic1}-\eqref{eq:ic2}.
At early times we expect that the growth can be described by
\begin{align}
D(t) \sim
\begin{cases}
D(0) e^{\Lambda t} & \textrm{for FPUT,} \\
D(0) t  & \textrm{for Toda.}
\end{cases}
\end{align}
However $\{\Delta{\bf q}(t), \Delta{\bf p}(t)\}$ cannot grow  forever since $\{{\bf q}(t),{\bf p}(t)\}$ are constrained to be on the constant energy surface and are themselves bounded. Let us define a ``phase-space covering'' time scale $\tau_{\rm ph}$ as the time at which each $\Delta{q}_j, \Delta{p}_j$ become of order $\sqrt{E/N}$ and  $D$ becomes of order $\sqrt{2E}$. Then we have
\begin{align}
\tau_{\rm ph} \sim
\begin{cases}
\Lambda^{-1} \ln \left( \frac{\sqrt{2E}}{D(0)}\right) & \textrm{for FPUT,} \\
\left( \frac{\sqrt{2E}}{D(0)}\right) & \textrm{for Toda.}
\end{cases}
\label{eq:chaos} 
\end{align}
The initial width $D(0)$ should be proportional to $\gamma$, which characterizes the width of our  initial phase-space distribution. The above arguments should work better for smaller $\gamma$. One  expects that $\tau_{\rm eq} \sim \tau_{\rm ph}$ and we now present some numerical results that support this. 
From Eq.~\eqref{eq:chaos}  we see that  $\tau_{\rm ph}$ and hence $\tau_{\rm eq}$ should scale with $\gamma$ as $\ln(\gamma)$ for the FPUT chain and as $1/\gamma$ for the Toda chain. 
In Fig.~(\ref{Fchaos}a) we see the logarithmic dependence of $\tau_{\rm eq}$ on  $\gamma$ for the FPUT chain, with $\alpha=0.0848$. In Fig.~(\ref{Fchaos}b) we show the dependence of $\tau_{\rm eq}$ on  $\gamma$ for the Toda chain, again with $\alpha=0.0848$. The slope on a log-log plot is close to $1$, supporting the expectation $\tau_{\rm eq} \sim 1/\gamma$. Next in Fig.~\eqref{tauvseps} we compare, for the FPUT system, the dependence of $\tau_{\rm ph}$ and $\tau_{\rm eq}$ on $\epsilon$ for $\gamma =10^{-8}, 0.9$. For $\gamma=10^{-8}$ we find $\tau_{\rm eq} \sim 1/\epsilon^{5.9}$ and $\tau_{\rm ph} \sim 1/\epsilon^{4.7}$. Thus we see a reasonable level of agreement though this is not perfect.

The system is translationally invariant. So, the results are identical if we initially distribute the energy to a different set of four successive particles, while maintaining the order of the initial  distribution. If the energies are distributed in a different permutation  we find that, while  the precise equilibration times are different, the dependence of $\tau_{\rm eq}$ on $\epsilon$ and $\gamma$ are still the same.

In Fig.~(\ref{Fchaos}a), the solid line describes an ensemble of initial conditions described by Eqs.~\eqref{eq:ic1}-\eqref{eq:ic2}, referred to as zero volume ensemble, since the latter occupies zero volume in the phase space. The magnitude of the slope of this line on a semilog plot is found to be $4.5\times 10^{3}$. Its inverse is $2.2\times 10^{-4}$, close to the Lyapunov exponent of the system $\Lambda\approx4.1\times 10^{-4}$. 
We have also studied an ensemble of initial conditions that has randomness in  all the degrees of freedom (while maintaining momentum conservation), thereby occupying a finite volume in the phase space,  quantified by the number $\gamma$. 
The results for this are shown by the dashed line in  Fig.~(\ref{Fchaos}a) where again we see the logarithmic dependence on $\gamma$ 
and in fact find a  closer agreement between the magnitude of the slope (the inverse of which is $3.3\times 10^{-4}$) and the Lyapunov exponent, thus making a stronger point for our claim regarding the relation between the equilibration of local observables and sensitive dependence of the system on initial conditions. 

We believe that the properties of the initial conditions, such as its symmetries and its vicinity from breather solutions can affect the exponential growth of the perturbations of the initial conditions, which would lead to a retardation of thermalization. Far away from breathers we can expect there is no such effect. This needs to be investigated further and is beyond the scope of this work. Nevertheless, this method gives us a way to link chaos and thermalization in the $\alpha-$FPUT system. We have also verified this relation for the $\beta$-FPUT system.

\section{Conclusions}
\label{sec:conclusions}
We studied the time scale of thermalization of local variables in the $\alpha$-FPUT chain and its two  limiting integrable versions, namely the harmonic chain and the Toda chain. Considering systems with $N=32$ particles and total energy $E=31$, we estimated the thermalization time $\tau_{\rm eq}$  by measuring $\langle z_i \partial H/\partial z_i \rangle$ ($z_i$ indicating phase space coordinates) and finding the time to attain equipartition. The averaging is done over initial conditions chosen from a distribution whose width is characterized by the parameter $0 \leq \gamma \leq 1$, with $\gamma=1$ corresponding to the broadest distribution and $\gamma=0$ corresponding to a fixed initial condition. The initial distribution is taken to be one where energy is localized initially in  real space instead of normal mode space.   
The system is described by a single dimensionless parameter $\epsilon=\alpha(E/N)^{1/2} $ characterizing the effective nonlinearity.  Some of our main findings are as follows:

(i) For the $\alpha$-FPUT chain we find  $\tau_{\rm eq} \propto {1}/{\epsilon^{a}}$ with $a$ between 4 and 6, and $\gamma$ dependent, in contrast to normal mode equilibration times  \cite{Onorato2015}, where one finds $a\approx8$.  

(ii) The thermalization time depends on the initial ensemble and we find $\tau_{\rm eq} \propto \ln(\gamma)$, with the proportionality constant being close to the inverse of the maximal Lyapunov exponent of the system, thus quantifying the relation between thermalization and chaos for the $\alpha$-FPUT system.

(iii) We find that local variables equilibrate at much shorter time scales for the normal mode localized initial conditions (NMLE) than phase space localized initial conditions (SLE).

(iv) Surprisingly, we find that the Toda chain equilibrated on very long time scales if the width of the initial distribution is broad enough. In fact we obtain $\tau_{\rm eq} \sim 1/\gamma$. On the other hand, the harmonic chain never equilibrates.

(v) We provide a simple geometric understanding of  these results -- the equilibration time is simply related to the time it takes for an ensemble of initial conditions in the $2N$ dimensional phase space to spread over the microcanonical energy surface. For the FPUT chain, for energies such that the system is chaotic with a positive Lyapunov exponent, a fast exponential (in time) spreading occurs. 
For the Toda chain the growth is linear and so thermalization takes more time.  We provide  numerical evidence to support  this picture. 

Thermalization in finite Hamiltonian systems has usually been studied by either considering a time averaging protocol or an ensemble averaging protocol. As we illustrate, they can lead to very different estimates for the time scale of equilibration. In our example, the time averaging protocol gives a thermalization time that is several orders of magnitude larger than that obtained from ensemble averaging. We believe that the ensemble averaging protocol is relevant for understanding aspects of the classical-quantum correspondence in the context of thermalization in finite systems. 
For quantum systems several studies show, e.g \cite{rigol2008}, that a finite quantum system prepared in a pure initial state and evolving under unitary dynamics can exhibit thermalization (without requiring any
time averaging). A corresponding statement for the classical system is difficult if (corresponding to the  quantum pure state) one considers a single point in phase space. However if we smear out the point into an initial blob in phase space, as is done in our study, then we can get an equivalent classical statement. The smearing out process can be thought of as arising from the uncertainty principle. As an example we point out recent work \cite{bukov2019} on the quantum-classical correspondence in Floquet systems where such an averaging protocol is essential.

As a concluding remark,  we note that another important question is that of thermalization of macroscopic systems. There we expect that thermalization requires neither  a temporal or an ensemble averaging protocol, but arises from the fact that physical observables are macro-variables and their measurement typically involves an averaging over many degrees of freedom \cite{Lebowitz1999,Goldstein2017}.

\section{Acknowledgments}
SG would like to thank Hemanta Kumar for help with computations  and Varun Dubey for useful discussions. The numerical simulations were done on \textit{Contra, Mario} and \textit{Tetris} computing clusters of ICTS-TIFR.

\bibliographystyle{unsrt}
\bibliography{references}

\end{document}